\documentclass[final,3p]{elsarticle}
\usepackage{amsfonts}
\usepackage{amssymb}
\usepackage{amsmath}
\usepackage{graphicx}

\usepackage[normalem]{ulem}
\usepackage{xcolor}

\begin{document}

\begin{frontmatter}

\title{Anisotropic Diffusion and Traveling Waves of Toxic Proteins in Neurodegenerative Diseases}

\author[umass,oxford]{P.G. Kevrekidis}

\author[oxford]{Travis Thompson}

\author[oxford]{Alain Goriely}

\address[umass]{Department of Mathematics and Statistics, University of Massachusetts Amherst,
Amherst, MA 01003-4515, USA}

\address[oxford]{Mathematical Institute, University of Oxford, Oxford, UK}

%\address[other]{Other Places}

\begin{abstract}
Neurodegenerative diseases are closely associated with the amplification and invasion of toxic proteins. In particular Alzheimer's disease is characterized by the systematic progression of  amyloid-$\beta$ and $\tau$-proteins in the brain.
These two protein families are coupled and it is believed that their joint presence greatly enhances the resulting damage.  Here, we examine a class of coupled chemical kinetics models of healthy and toxic proteins in two
spatial dimensions.  The anisotropic diffusion expected to
take place within the
brain along axonal pathways
is  factored in the models and produces a filamentary, predominantly
one-dimensional transmission. Nevertheless, the potential of
the anisotropic  models towards generating interactions taking advantage of
the two-dimensional landscape is showcased. Finally, a reduction
of the models into a simpler family of generalized
Fisher-Kolmogorov-Petrovskii-Piskunov (FKPP) type systems is examined.
It is seen that the latter captures well the qualitative propagation
features, although it may somewhat underestimate the concentrations of
the toxic proteins.
\end{abstract}

\begin{keyword}
Alzheimer disease, brain, traveling waves, reaction-diffusion equations, Amyloid-$\beta$,
$\tau$-Proteins,
Chemical kinetics, FKPP models. 
\end{keyword}

\end{frontmatter}

\section{Introduction}\label{model}

Neurodegenerative disorders are both particularly complex from a scientific and clinical point of view and
especially costly from a human and economic perspective. The ongoing
effort to address them involves an extensive pipeline of drug
development (see, e.g.,~\cite{cummings}) with very poor success. 
It is therefore crucial to enhance our fundamental understanding of the development
of such disorders. One of the prevalent viewpoints involves the
critical relevance of the formation of protein aggregates, referred to as prion-like aggregates by similarity to prion diseases~\cite{prlr1}. In this approach certain proteins develop a toxic variant that 
can be thought of as seeding the relevant
``infection'', progressively leading to an autocatalytic chain reaction of misfolded
and aggregated toxic proteins that, in turn, grow and spread
throughout the brain inhibiting proper cell function.
This type of toxic disruption, unless halted or removed, continues
to  disrupt progressively the nervous system, ultimately leading
to atrophy of different parts of the brain, causing the degradation and eventual death of the
patient~\cite{prlr3}.

An interesting aspect of the prion-like hypothesis is that these proteins
are different in different diseases and seeded at different locations, yet there are some universal features of the process and the resulting biomarkers follow similar trends. For example, in Alzheimer's disease,
the relevant proteins have been recognized to be  amyloid-$\beta$
(A$\beta$) and  $\tau$-protein ($\tau$P). Intriguingly, the two differ significantly in the way they aggregate, their location in the brain and where they originate: the former forms extracellular
aggregates and plaques, while the latter operates intra-cellularly,
cross-linking microtubules and inducing the formation of large
disorganized neurofibrillary tangles~\cite{conr1,conr2}.
A similar prion-like growth has been argued to be of relevance
to the cases of Parkinson's disease with $\alpha$-synuclein playing a similar role  and
in amyotrophic lateral sclerosis where the principal biomarker is the TAR DNA binding protein,
TDP-43~\cite{ref1,ref2,prlr4,prlr5}.

Here, we will  focus on Alzheimer's disease (AD) and the associated dynamics of  A$\beta$ and $\tau$P.
The ubiquitous presence of A$\beta$ plaque in AD patients lead to the so-called ``amyloid cascade
hypothesis'' of Hardy and collaborators~\cite{harr1,harr2}, which has
contributed significantly to the research directions of the relevant
field for over 25 years~\cite{harr3}. Indeed, at the present stage
of drug development~\cite{cummings}, about half of the current efforts
towards disease-modifying therapies, and about 32\% of the total
of currently tested drugs, are primarily focused on
A$\beta$ and are classified as ``anti-amyloid'' drugs. Also,
another 4\% of currently tested drugs are ``anti-$\tau$'' and yet another 4\% are both
anti-amyloid and anti-$\tau$. In total, a remarkable 40\% of all currently
developed drugs focus exclusively on A$\beta$ and $\tau$P, while only 21\% focus on other disease-modifying factors.
For reference, another 28\% of potential drugs targets neuropsychiatric symptoms, while 11\% focuses on cognitive enhancers.

These efforts have also motivated attempts to 
model the progression of A$\beta$ and $\tau$P toxicity~\cite{g1,g2,al1} 
in line with corresponding measurements of such biomarkers
from different types of scans~\cite{bateman}.
Indeed, recent efforts have shown that at least some of the clinical
antibodies under consideration, such as aducanumab, may hold
promise towards binding with fibrillar forms of A$\beta$,
reducing the flux of its oligomeric forms and ultimately yield
positive outcomes in clinical trials~\cite{linse}.

Our aim in the present work is to present some simplifying models
of the coupled dynamics of A$\beta$ and $\tau$P in the spirit
of~\cite{g1,g2,al1},
but focusing predominantly in two spatial dimensions as it has been shown to capture most of the qualitative dynamics for the progression of a single toxic protein \cite{g2}. 
%While this is a clearly simpler setting than the full-3d
%geometry of the brain, it still offers a perspective on the effects of
%geometry, as well as anisotropy, while incorporating the nonlinear
%interactions between the proteins. As such, while a definitive
%simplification,
%it can be thought of as a reasonable, less computationally expensive compromise that
%incorporates
%some of the important biological and physico-chemical ingredients of the system.
Our spatial-temporal model is originally based on chemical
kinetics in line with the recent efforts of~\cite{g2,al1}, rather than
starting from a purely phenomenological Fisher-KPP (FKPP for short, named after the ubiquituous
Fisher-Kolmogorov-Petrovskii-Piskunov model)~\cite{murray}, as e.g., in~\cite{g1}.
%; we will refer to the latter hereafter as FKPP for short. 
It incorporates 4 populations, namely healthy and toxic variants
of the A$\beta$ and $\tau$P. In addition to the generation of the
healthy proteins and the decay of both healthy and toxic ones with
corresponding
rates, the conversion of healthy A$\beta$ and $\tau$P into respective
toxic variants is accounted for and so is the role of toxic A$\beta$ towards
catalyzing the production of toxic $\tau$P~\cite{g2,al1}. We observe
different
scenarios of so-called primary tauopathies (where each toxic species
can exist on its own and they can also co-exist) and so-called
secondary
tauopathies, where the existence of toxic $\tau$P is contingent
upon the presence of toxic A$\beta$. We examine each scenario in a
quasi-1d setting where the role of the second dimension is simply in
providing  a weak lateral spreading of the waves. We also examine
a genuinely 2d scenario, where the transverse interaction
of
the toxic waves is critical for the spreading of the disorder, as a
way of
illustrating the effect of dimensionality. Finally, we illustrate how
to approximate the model by an effective (generalized) FKPP variant
and compare the latter with the full results finding very good qualitative
agreement despite the partial quantitative disparities between the
two.

Our presentation is structured as follows. In section II, we discuss
the
models at the different levels of description (chemical kinetic
4-component
model and its two-component FKPP reduction) and some of their
salient features. Then, in section III, we will present the different
simulations
for primary and secondary tauopathies, without and with incorporating
a significant role of the transverse degrees of freedom. The
comparison
to the FKPP  will also be given to obtain
a sense of the relevance of studying the less computationally
expensive models of the latter type. Finally, in section IV, we will
summarize our findings and present our conclusions.

\section{Model Formulation}

Our principal model, presented in \cite{al1}, features four species: healthy A$\beta$ and $\tau$P
and toxic A$\beta$ and $\tau$P. Here are the processes involved
in each species:

\begin{itemize}

\item Healthy A$\beta$ diffuses anisotropically over the flat domain.
  It is produced with a constant rate $a_0$ and is cleared with
  a clearance rate $a_1$. Moreover, the interaction of healthy and
  toxic A$\beta$ results in the toxification of healthy A$\beta$
  proteins. The healthy A$\beta$ concentration is denoted as $u(x,y,t)$,
  with the independent variables being $(x,y)$ in space and $t$ for time.

\item In a similar vein, toxic A$\beta$ proteins diffuse with similar
  diffusivities anisotropically over our two-dimensional  slice.
  They are cleared with a rate $\tilde{a}_1$ and get produced by
  the toxification of the corresponding healthy concentration.
  With tildes being used for the toxic species, the relevant
  population is denoted by $\tilde{u}(x,y,t)$.

\item Similarly, the healthy $\tau$P concentration is denoted by
  $v(x,y,t)$ and involves diffusion, production at a constant rate
  $b_0$ and clearance with a clearance rate $b_1$. Here, toxic $\tau$P
  are produced either by direct interaction of healthy and toxic
  $\tau$P with rate $b_2$
  or catalyzed by the (surrounding) presence of toxic A$\beta$ with
  rate
  $b_3$.

\item Finally, similar anisotropic diffusion properties are posited
  also for toxic $\tau$P, with a clearance rate $\tilde{b}_1$
  and production by the two above mechanisms of toxification of
  the healthy $\tau$P population.

\end{itemize}

Mathematically translating the above 4 populations and the respective
assumptions, we obtain the following nonlinear partial differential
equations:
\begin{eqnarray}
  u_t&=& D_x \left(u_{xx} + \epsilon u_{yy}\right) + a_0 -a_1 u - a_2 u
  \tilde{u},
         \label{eqn1}
  \\
  \tilde{u}_t &=& D_x \left( \tilde{u}_{xx} + \epsilon
                  \tilde{u}_{yy}\right)
                  -\tilde{a}_1 \tilde{u} + a_2 u \tilde{u},
                  \label{eqn2}
  \\
  v_t &=& D_x \left(v_{xx} + \epsilon v_{yy}\right) + b_0 -b_1 v - b_2 v
          \tilde{v} - b_3 v \tilde{u} \tilde{v},
          \label{eqn3}
  \\
  \tilde{v}_t &=& D_x \left( \tilde{v}_{xx} + \epsilon
                  \tilde{v}_{yy}\right)
                  - \tilde{b}_1 \tilde{v} + b_2 v
          \tilde{v} + b_3 v \tilde{u} \tilde{v}.
          \label{eqn4}
\end{eqnarray}
Here, for simplicity we have assumed that all the diffusivities are
equal
and are assigned to be $D_x$ along the $x$-axis, while they are
$D_y=\epsilon D_x$ along the $y$-axis. All parameters and variables are assumed to be positive. Finally, the subscripts denote
partial derivatives with respect to the corresponding independent
variables.

Following also the considerations of~\cite{al1}, one  defines
a ``damage'' variable $q(x,y,t)$ based on the following (trivial in
space) PDE:
\begin{eqnarray}
  q_t= \left(k_1 \tilde{u} + k_2 \tilde{v} + k_3 \tilde{u}
  \tilde{v}\right) (1-q).
  \label{eqn5}
\end{eqnarray}
This naturally tends to a stable fixed point of $q(x,y,t)=1$ (maximal damage), starting
from an initial condition of no-damage, i.e., $q(x,y,0)=0$.

A relevant consideration is that of identifying the fixed points in
this model. There are, generally speaking, 4 equilibrium fixed points
in this case.
\begin{enumerate}
\item $(\frac{a_0}{a_1},0,\frac{b_0}{b_1},0)$ is the always unstable
  (in the realm of this model) healthy state. The assumption here is
  that we are modeling an early stage of the emergence of the
  neurodegenerative
  disorder.
  \item
    $(\frac{\tilde{a}_1}{a_2},\frac{a_0}{\tilde{a}_1}-\frac{a_1}{a_2},\frac{b_0}{b_1},0)$
    is a state devoid of toxic $\tau$P, but bearing toxic A$\beta$.
    For this state to be biologically meaningful (i.e., reflecting
    positive concentrations), the assumption is  $a_0 a_2> a_1 \tilde{a}_1$.
    \item Similarly,
      $(\frac{a_0}{a_1},0,\frac{\tilde{b}_1}{b_2},\frac{b_0}{\tilde{b}_1}-
        \frac{b_1}{b_2})$ is a state with only healthy A$\beta$, but
        bearing both healthy and toxic $\tau$P. Here, biological
        relevance dictates that $b_0 b_2> b_1 \tilde{b}_1$.
      \item Lastly, there exists a homogeneous state with
        all four populations, healthy and toxic ones alike, being
        non-vanishing,
        whereby $u=\tilde{a}_1/a_2$,
        $\tilde{u}=\frac{a_0}{\tilde{a}_1}-\frac{a_1}{a_2}$,
        $v=\frac{\tilde{a}_1 a_2 \tilde{b}_1}{P}$ and
        $\tilde{v}=\frac{b_0}{\tilde{b}_1}- \frac{\tilde{a}_1 a_2
          b_1}{P}$,
        where $P=\tilde{a}_1 a_2 b_2 - a_1 \tilde{a}_1 b_3 + a_0 a_2
        b_3$.
        For all 4 equilibria to be present, it is necessary that
        {\it both} inequality constraints are satisfied enabling the previous two equilibria
        to exist. Interestingly, in this setting, the concentration of
        the
        toxic A$\beta$ (at equilibrium) remains the same as for the equilibrium devoid
        of toxic $\tau$P, yet the concentration of toxic $\tau$P is higher than that in
        which the only toxic species is $\tau$P. This property is a consequence of the one-way coupling (A$\beta$ influences the production of $\tau$P but $\tau$P does not influence A$\beta$, as observed experimentally).
        \end{enumerate}

The above observation leads to a classification of the so-called
tauopathies. By this, we will mean scenarios involving toxic
contributions
from both  A$\beta$ and $\tau$P. In the case of a \textit{primary tauopathy}, both of the above inequalities are
satisfied, then all 4 equilibria will exist. For a  \textit{secondary tauopathy}, we have $a_0 a_2> a_1
\tilde{a}_1$,
while $b_0 b_2<  b_1 \tilde{b}_1$, it is still possible to
have an equilibrium where both toxic components are concurrently
present, yet $\tau$P cannot be toxic by itself (i.e., in the absence of
toxic A$\beta$). Naturally for this scenario of secondary tauopathy to
occur, the relevant coefficient $b_3$ should be sufficiently large.
%(the relevant inequality for positivity of $\tilde{v}$ can be
%straightforwardly
%extracted, yet given its tedious form, we do not provide it here).
We will examine both of these scenarios in what follows.

Lastly, we consider the reduction of the model into a pair of FKPP-type
PDEs for the toxic components alone. To do so, an effective assumption
of
sufficiently larger (than the toxic) healthy concentrations of the two proteins is
relevant to incorporate. In particular, assuming an effectively
space- and time-independent
concentration of healthy A$\beta$ yields $u=a_0/(a_1 + a_2
\tilde{u})$.
This, in turn, under these assumptions of $\tilde{u} \ll u$ can be
approximated by $u \approx \frac{a_0}{a_1} (1- \frac{a_2}{a_1}
\tilde{u})$.
In a similar vein, we can extract, via leading order Taylor expansion,
$v=\frac{b_0}{b_1} (1- \frac{b_2}{b_1} \tilde{v} - \frac{b_3}{b_1}
\tilde{u} \tilde{v})$. 
Then, the resulting generalized FKPP equations stemming from the
substitution
of these approximations into Eqs.~(\ref{eqn2}) and~(\ref{eqn4}) are:
\begin{eqnarray}
  \tilde{u}_t &=& D_x \left( \tilde{u}_{xx} + \epsilon
                  \tilde{u}_{yy}\right)
                  + \left( \frac{a_2 a_0}{a_1} - \tilde{a}_1\right)
                  \tilde{u}
                  - \frac{a_2^2 a_0}{a_1^2} \tilde{u}^2,
                  \label{eqn2b}
  \\
   \tilde{v}_t &=& D_x \left( \tilde{v}_{xx} + \epsilon
                  \tilde{v}_{yy}\right)
                  + \left( \frac{b_2 b_0}{b_1}-\tilde{b}_1+ \frac{b_3
                   b_0}{b_1} \tilde{u}\right) \tilde{v} -
                   \left( \frac{b_2^2b_0}{b_1^2} + \frac{2 b_2 b_3
                   b_0}{b_1^2} \tilde{u} + \frac{b_3
                   b_0}{b_1^2}\right) \tilde{v}^2.
          \label{eqn4b}
\end{eqnarray}
We will also explore the results of the system of
Eqs.~(\ref{eqn2b})-(\ref{eqn4b}) and compare it with the observations
stemming from Eqs.~(\ref{eqn1})-(\ref{eqn4}), as concerns the
evolution
of both primary and secondary tauopathies in what follows. 
Linearized theory predicts the speeds of propagation of the
corresponding resulting fronts, namely for the front interpolating
between states $1$ and $2$, we have:
\begin{eqnarray}
  c^{12}=2 \sqrt{D_x \left( \frac{a_2 a_0}{a_1}- \tilde{a}_1\right)}.
  \label{en6}
\end{eqnarray}
We consider here the speed of propagation along the dominant direction
of diffusion, namely the x-axis, since we will assume $\epsilon \ll 1$
in what follows.
On the other hand, for the front interpolating between the homogeneous states
$1$ and $3$, we will have, respectively:
\begin{eqnarray}
c^{13}=2 \sqrt{D_x \left( \frac{b_2 b_0}{b_1}- \tilde{b}_1\right)}.
\label{eqn7}
\end{eqnarray}
Once the right propagating wave of the left blob and the left one of
the
right blob reach each other and interact, they will achieve a state
of co-existence and the resulting propagation speed that is obtained
via linearization around the co-existence state is:
\begin{equation}
c^{24}=
2 \sqrt{\frac{\tilde{\rho}_2}{a_2b_1 \tilde{a}_1}} %
\sqrt{\tilde{a}_1 \left(a_2 \left(b_0
   b_2-b_1 \tilde{b}_1\right)-a_1 b_0 b_3\right)+a_0 a_2 b_0
   b_3}.
\end{equation}
Having set up the relevant models,
we now turn to the corresponding numerical results.

\section{Numerical Results}

\subsection{Primary Tauopathy}

\begin{figure}[htbp]
\centering
\includegraphics[scale=0.25]{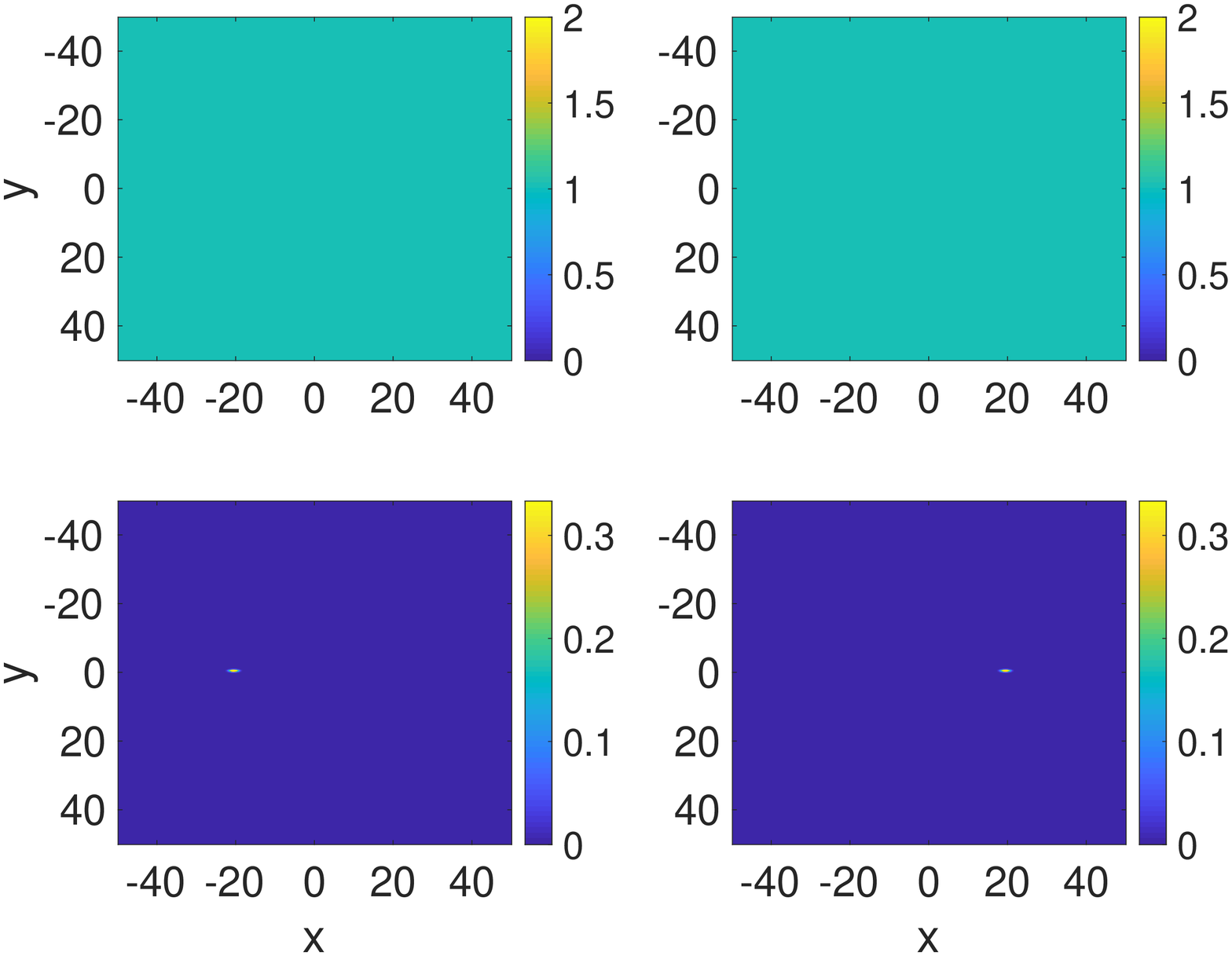}
\includegraphics[scale=0.25]{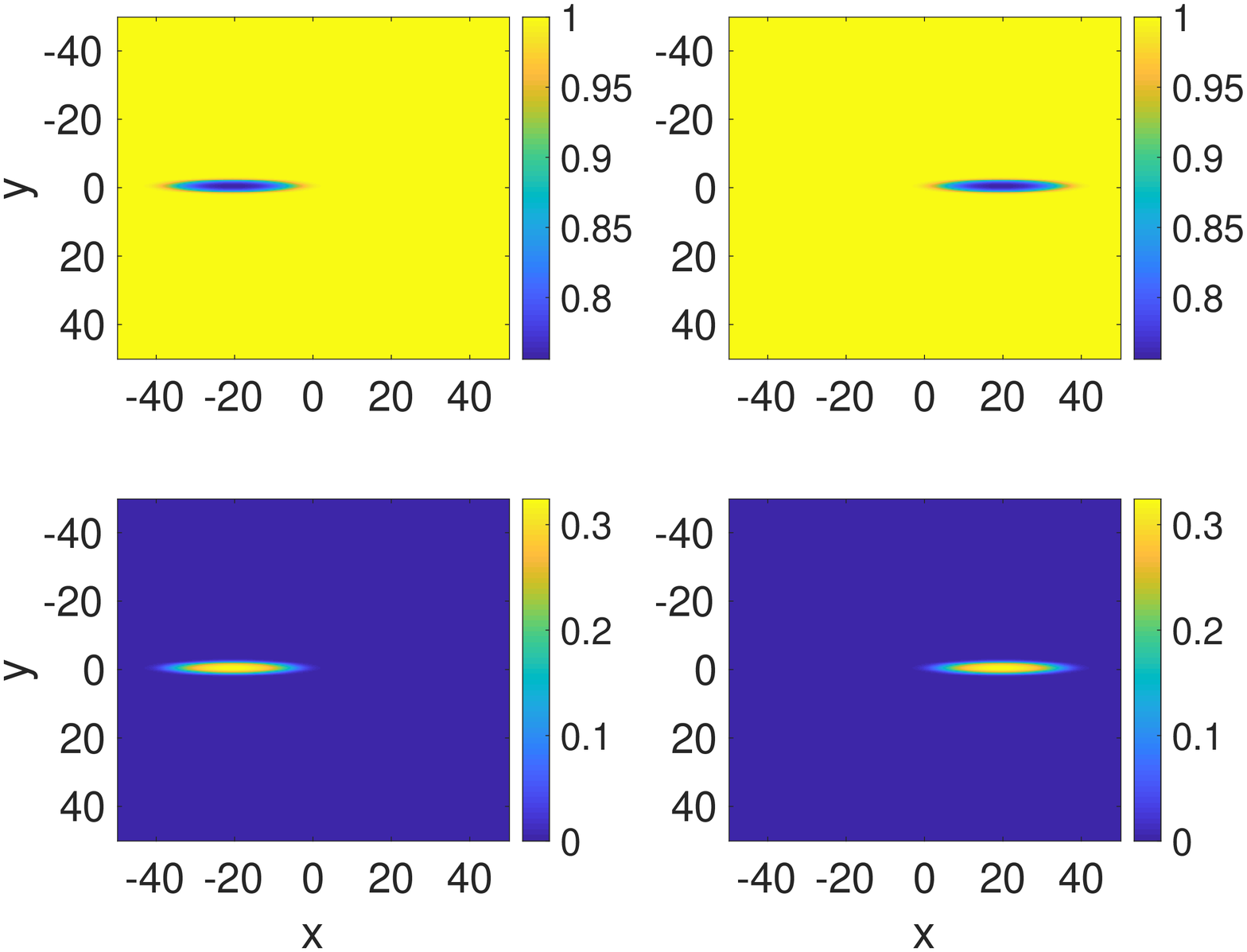}   \\
\includegraphics[scale=0.25]{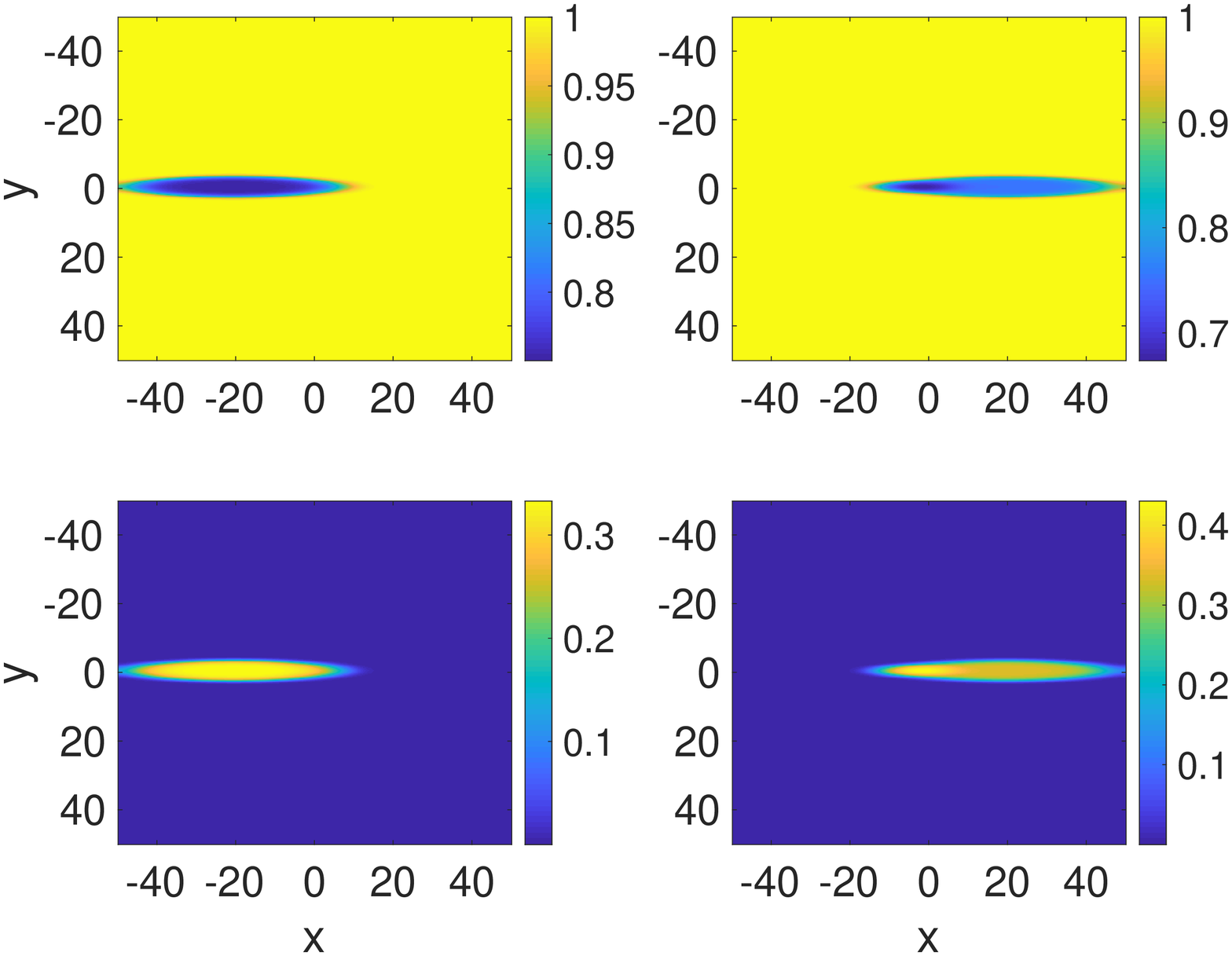}
\includegraphics[scale=0.25]{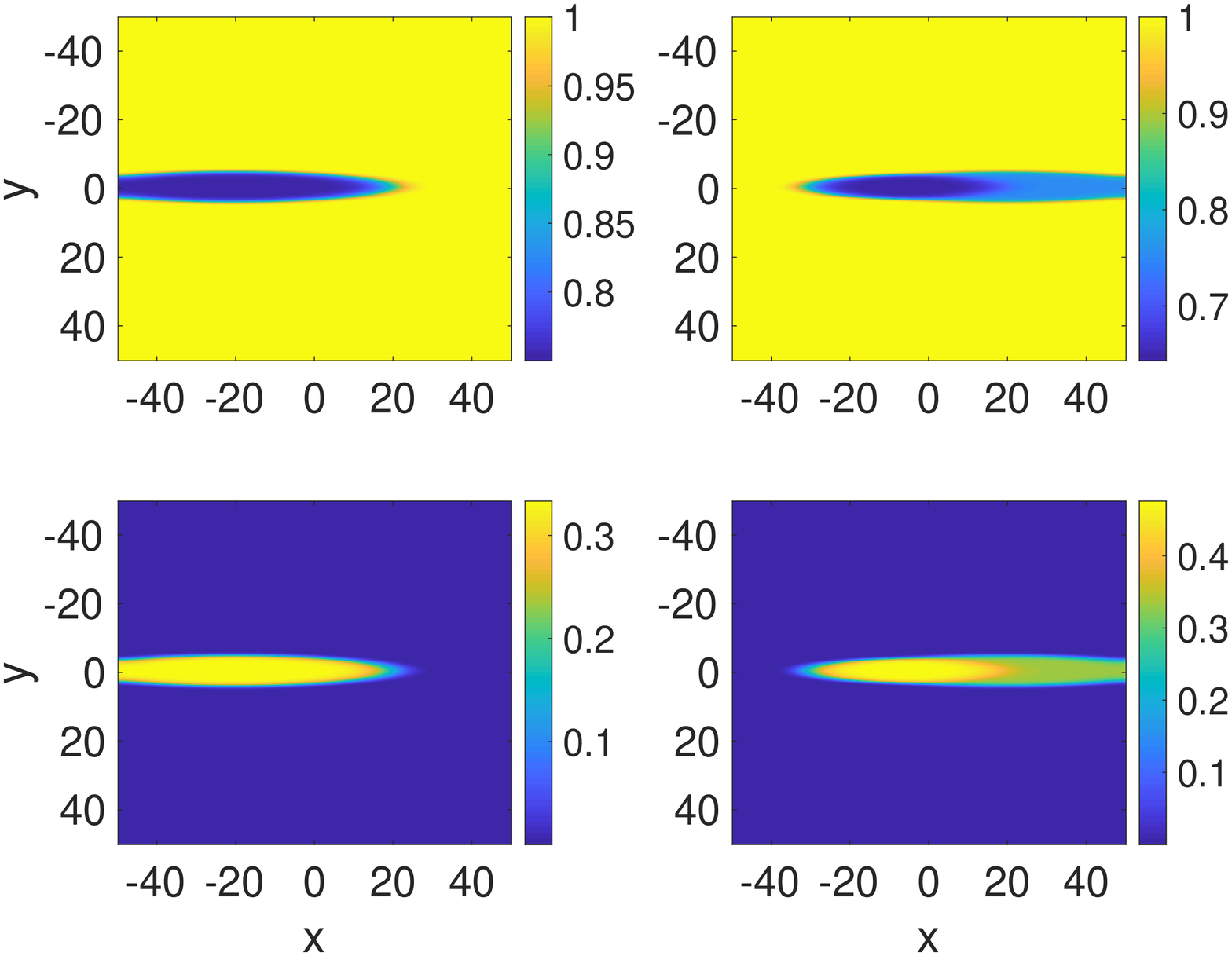}
\caption{Numerical evolution snapshots of a primary tauopathy via 4
  $2 \times 2$ sets of 4 panels each.
  The first set of panels is at $t=0.5$, the second at $t=30$, the third
  at $t=50$ and the last at $t=70$. What is shown is a contour plot of
the spatial distribution of all four of the relevant field concentrations: the
healthy A$\beta$ (top left), the healthy $\tau$P (top right),
the toxic A$\beta$ (bottom left) and the toxic $\tau$P (bottom right).}
\label{dark_sp}
\end{figure}

We start our exposition of the numerical results by examining a
setting of primary tauopathy (i.e., where all 4 relevant uniform
equilibrium
states exist). In this setting the 2nd state (involving no toxic
$\tau$P) and the 3rd state (involving no toxic A$\beta$) are only
attracting in the {\it absence} of one of the toxic species.
When both toxic species are present, the situation favors
the co-existing state where both toxic species are present (i.e., the
4th one). Hence, we design the following numerical experiment:
on the one side, we seed a narrow blob of toxic A$\beta$, while
on the other side, we seed a similar blob but of toxic $\tau$P, so as to see
how the respective toxicities will interact upon their propagation.
%Indeed, linearized theory predicts the speeds of propagation of the
%corresponding resulting fronts, namely for the front interpolating
%between states $1$ and $2$, we have:
%\begin{eqnarray}
%  c^{12}=2 \sqrt{D_x \left( \frac{a_2 a_0}{a_1}- \tilde{a}_1\right)}.
%  \label{en6}
%\end{eqnarray}
%On the other hand, for the front interpolating between the homogeneous states
%$1$ and $3$, we will have, respectively:
%\begin{eqnarray}
%c^{13}=2 \sqrt{D_x \left( \frac{b_2 b_0}{b_1}- \tilde{b}_1\right)}.
%\label{eqn7}
%\end{eqnarray}
%Once the right propagating wave of the left blob and the left one of
%the
%right blob reach each other and interact, they will achieve a state
%of co-existence and the resulting propagation speed that is obtained
%via linearization around the co-existence state
%\begin{equation}
%c^{24}=
%2 \sqrt{\frac{\tilde{\rho}_2}{a_2b_1 \tilde{a}_1}} %
%\sqrt{\tilde{a}_1 \left(a_2 \left(b_0
%   b_2-b_1 \tilde{b}_1\right)-a_1 b_0 b_3\right)+a_0 a_2 b_0
%   b_3}.
%\end{equation}
%unox=1/3*(sech((XX+20).^2+10*(ZZ-2.5).^2));
%vnox=1/3*(sech((XX-20).^2+5*(ZZ+2.5).^2));
In this primary tauopathy, we select $a_0=b_0=a_1=a_2=b_1=b_2=1$,
while $\tilde{a}_1=\tilde{b}_1=3/4$ and $b_3=1/2$.
The initial conditions associated with this numerical experiment shown in
Fig.~\ref{dark_sp} involve uniform profiles $u(x,y,0)=1$, $v(x,y,0)=1$
for healthy A$\beta$ and $\tau$P, while
for the toxic proteins we assume a small blob of initial
concentrations
in the form:
\begin{eqnarray}
  \tilde{u}(x,y,0)&=&\frac{1}{3} {\rm sech}^2\left((x+20)^2 + 10 y^2\right),
  \label{in1}
  \\
  \tilde{v}(x,y,0)&=&\frac{1}{3} {\rm sech}^2\left((x-20)^2 + 10 y^2\right).
                      \label{in2}
\end{eqnarray}
Notice that the relevant results have been found to be generic within
their
corresponding regimes of parametric inequalities, hence the
particular value of the parameters, as well as the amplitude and
precise
shape of the initial condition blobs do not play a crucial role as
regards the phenomenology reported below.

\begin{figure}[htbp]
\centering
\includegraphics[scale=0.25]{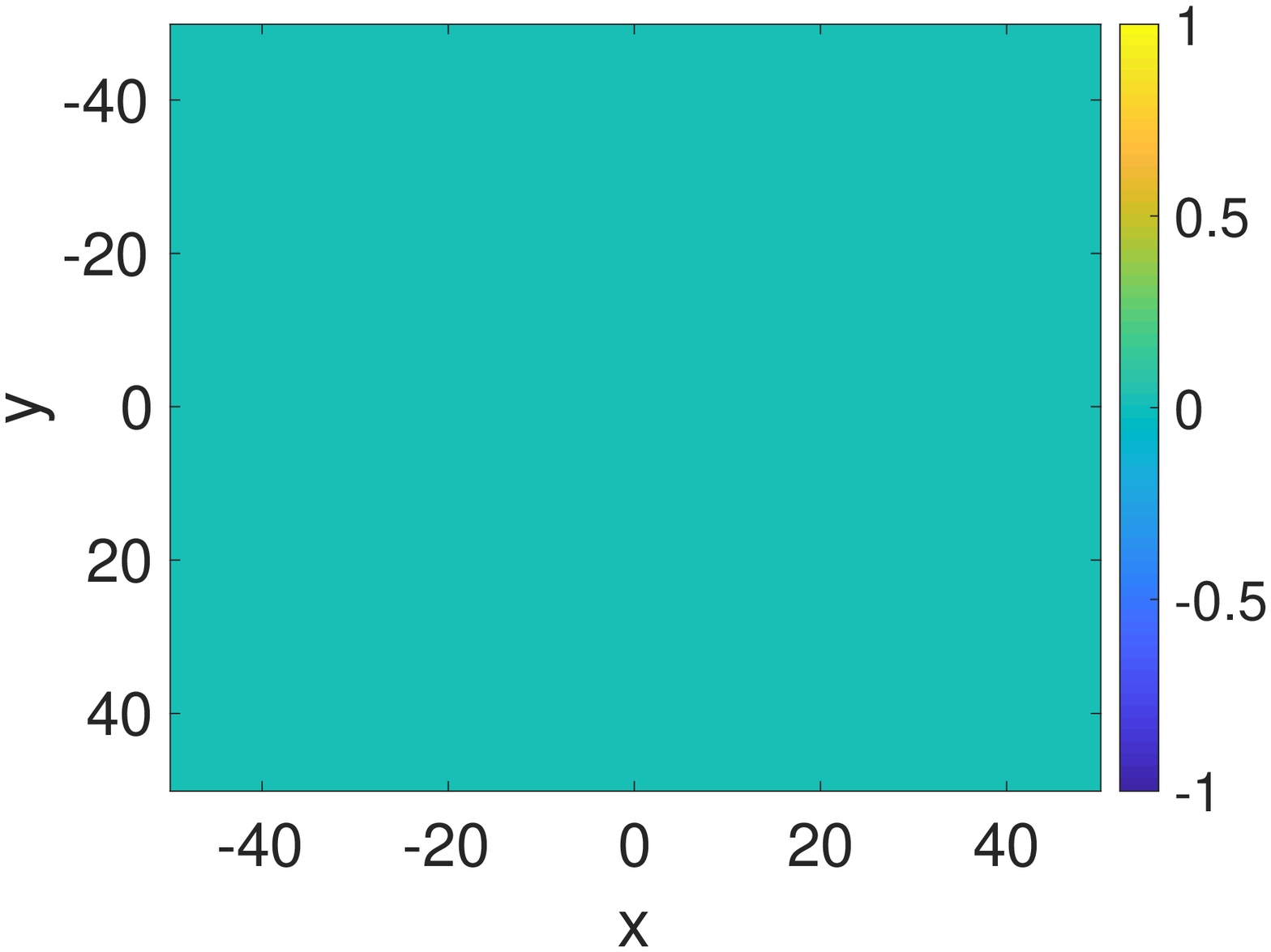}
\includegraphics[scale=0.25]{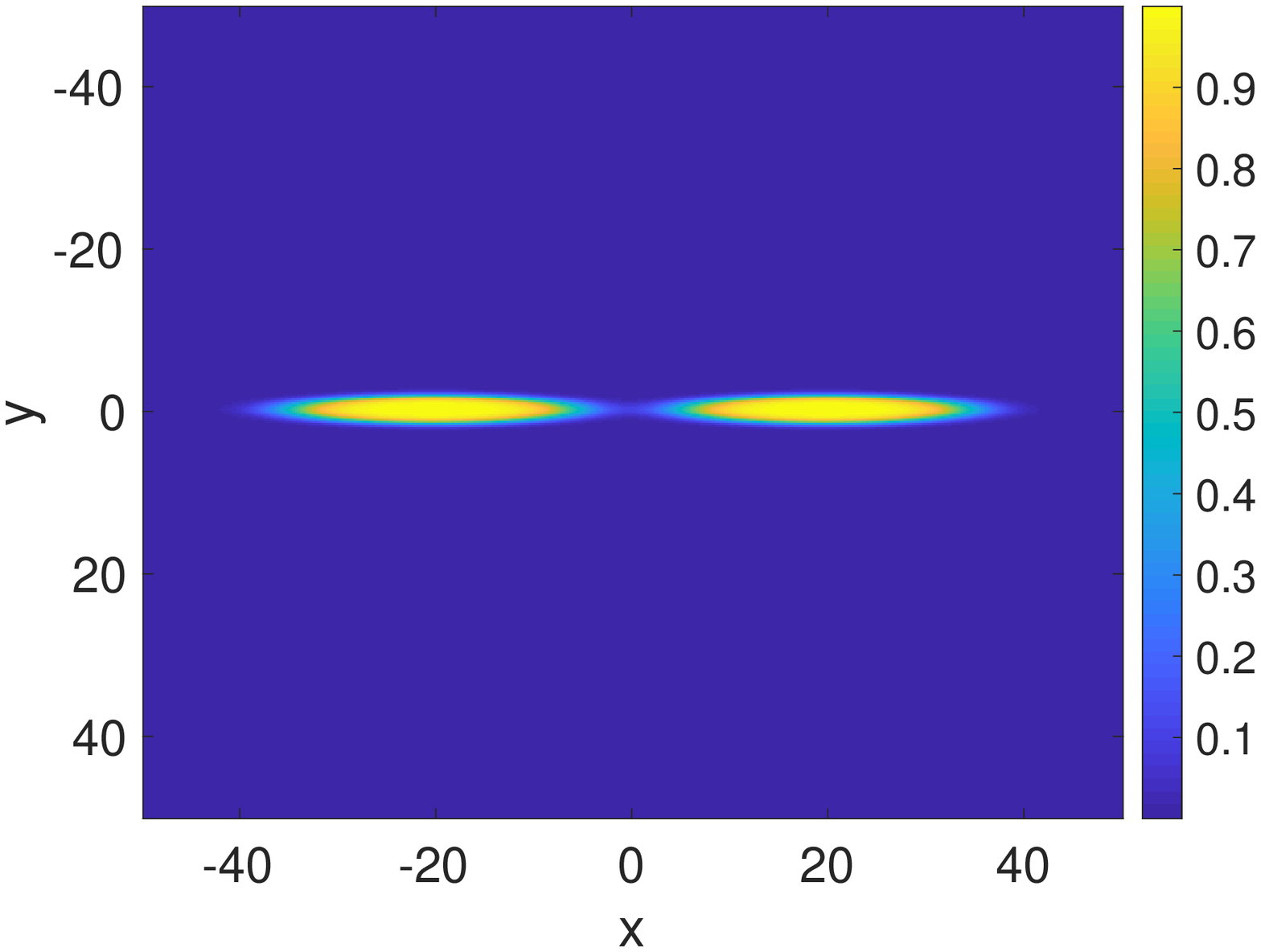}   \\
\includegraphics[scale=0.25]{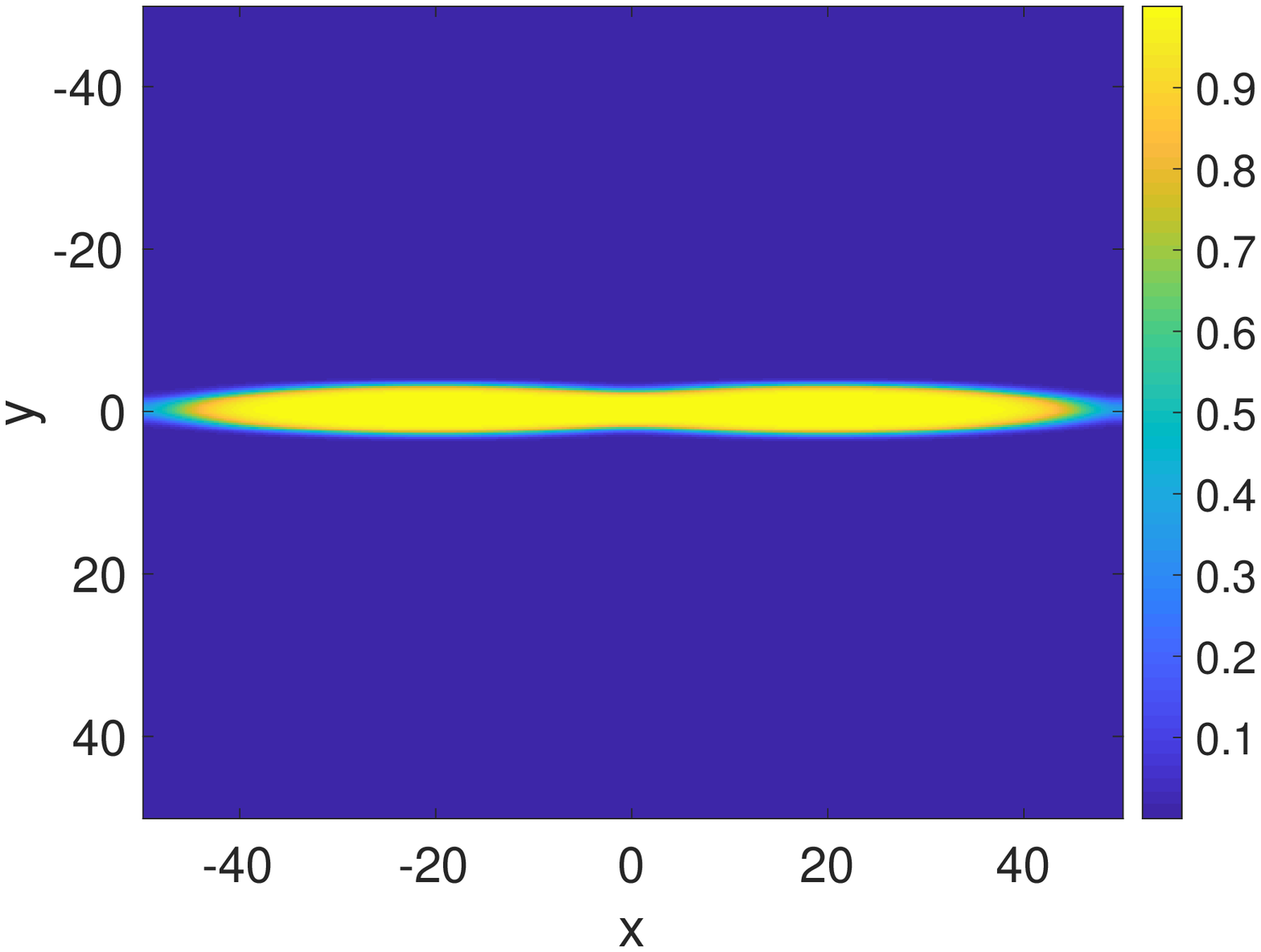}
\includegraphics[scale=0.25]{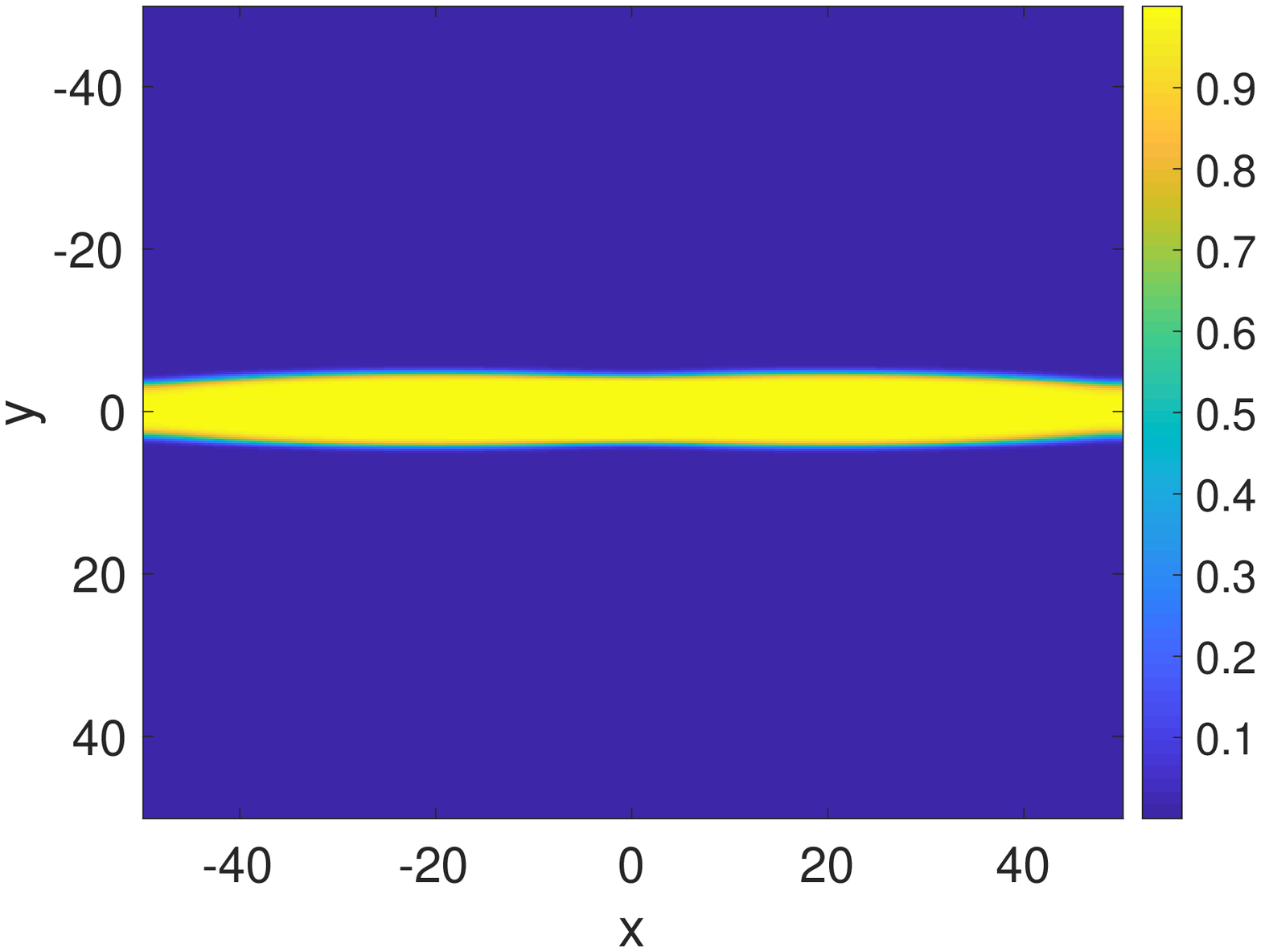}
\caption{Evolution of the damage function $q(x,y,t)$ at the same
  times as for the above simulation. I.e., $t=0.5$ at the top left,
  $t=30$ at the top right, $t=50$ at the bottom left and $t=70$ at the
  bottom right. It can be clearly seen how the evolving initial spots
  expand
into a ``corridor'' of damage over the dynamical evolution.}
\label{dark_sp2}
\end{figure}

\begin{figure}[htbp]
\centering
\includegraphics[scale=0.25]{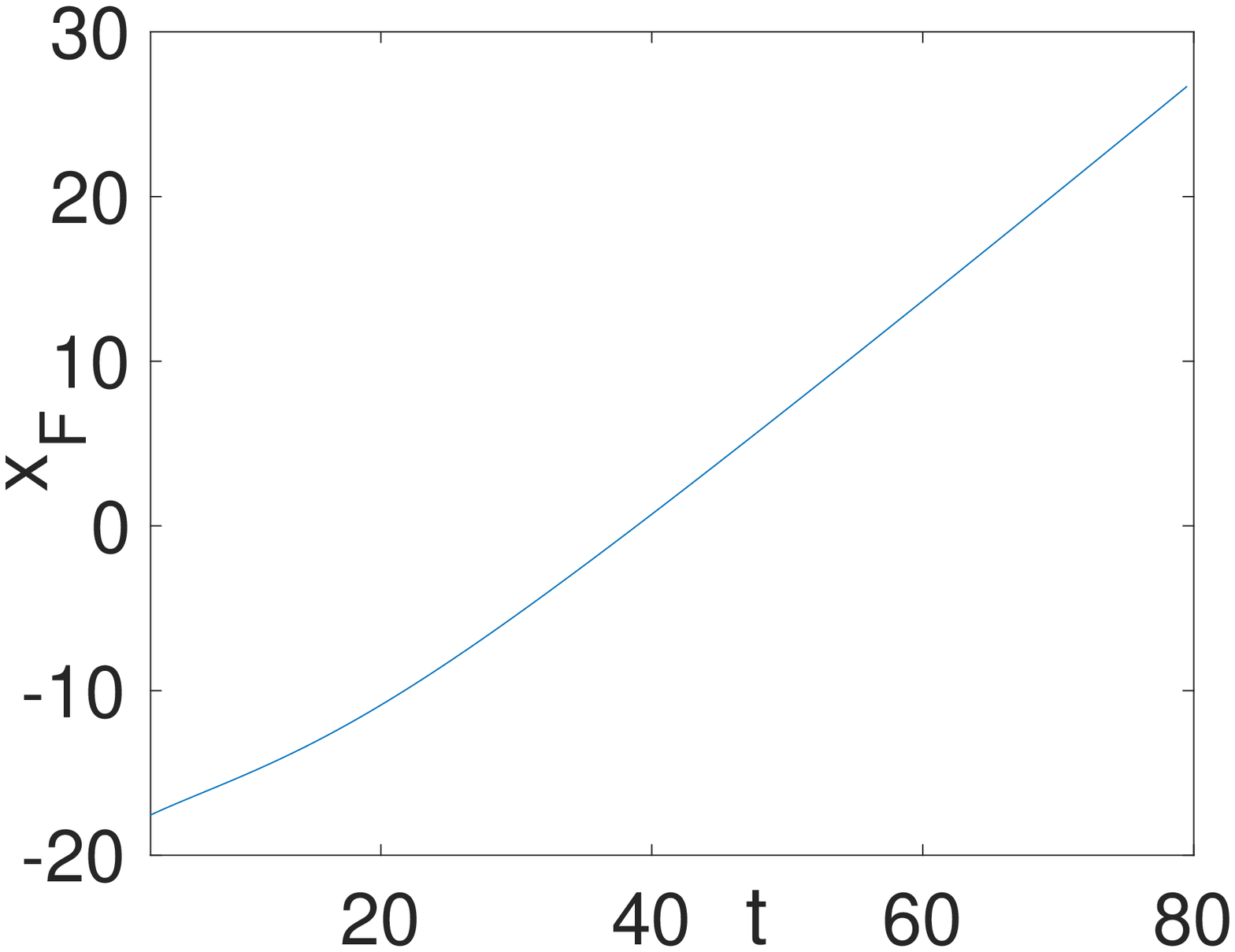}
\includegraphics[scale=0.25]{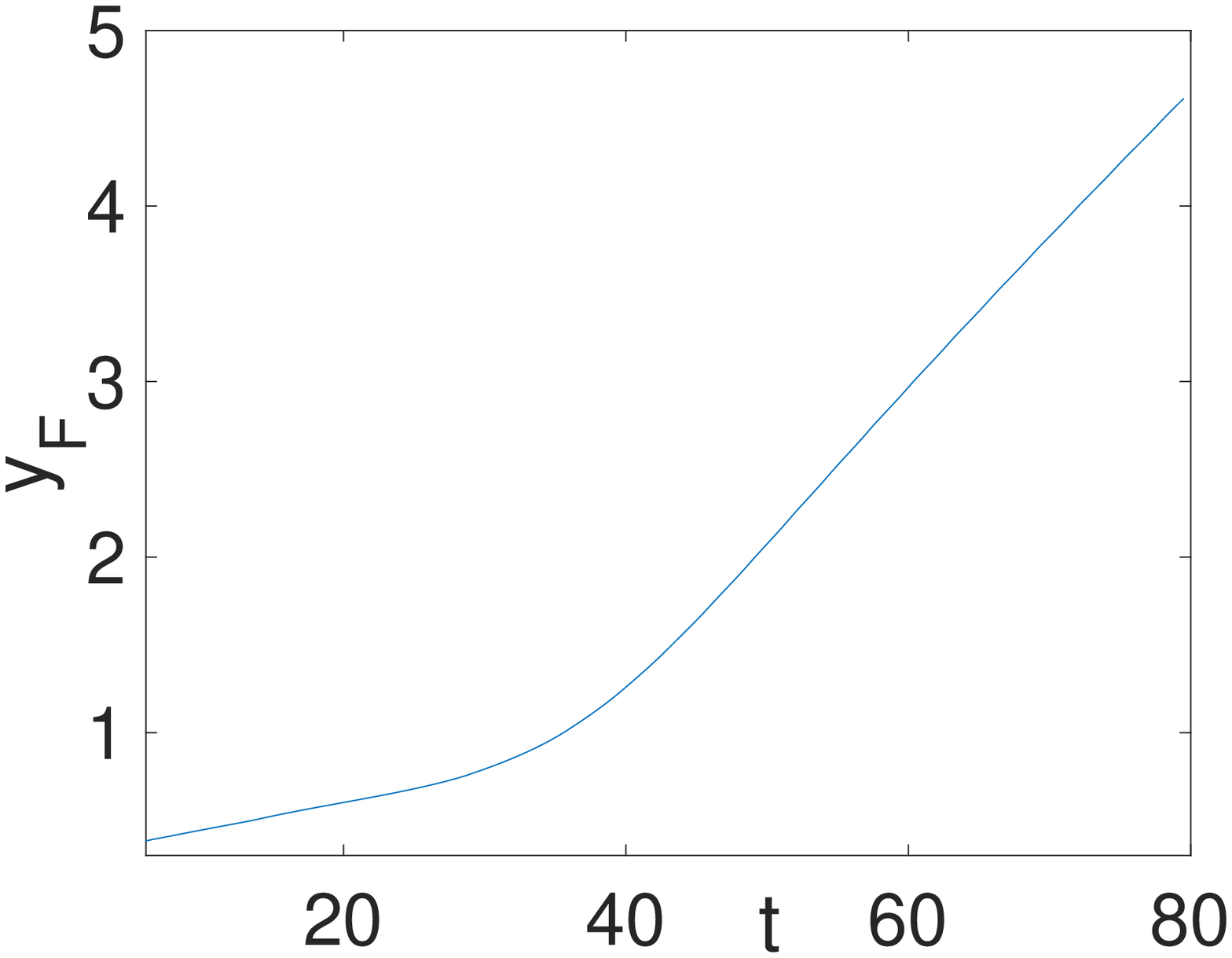}
\includegraphics[scale=0.25]{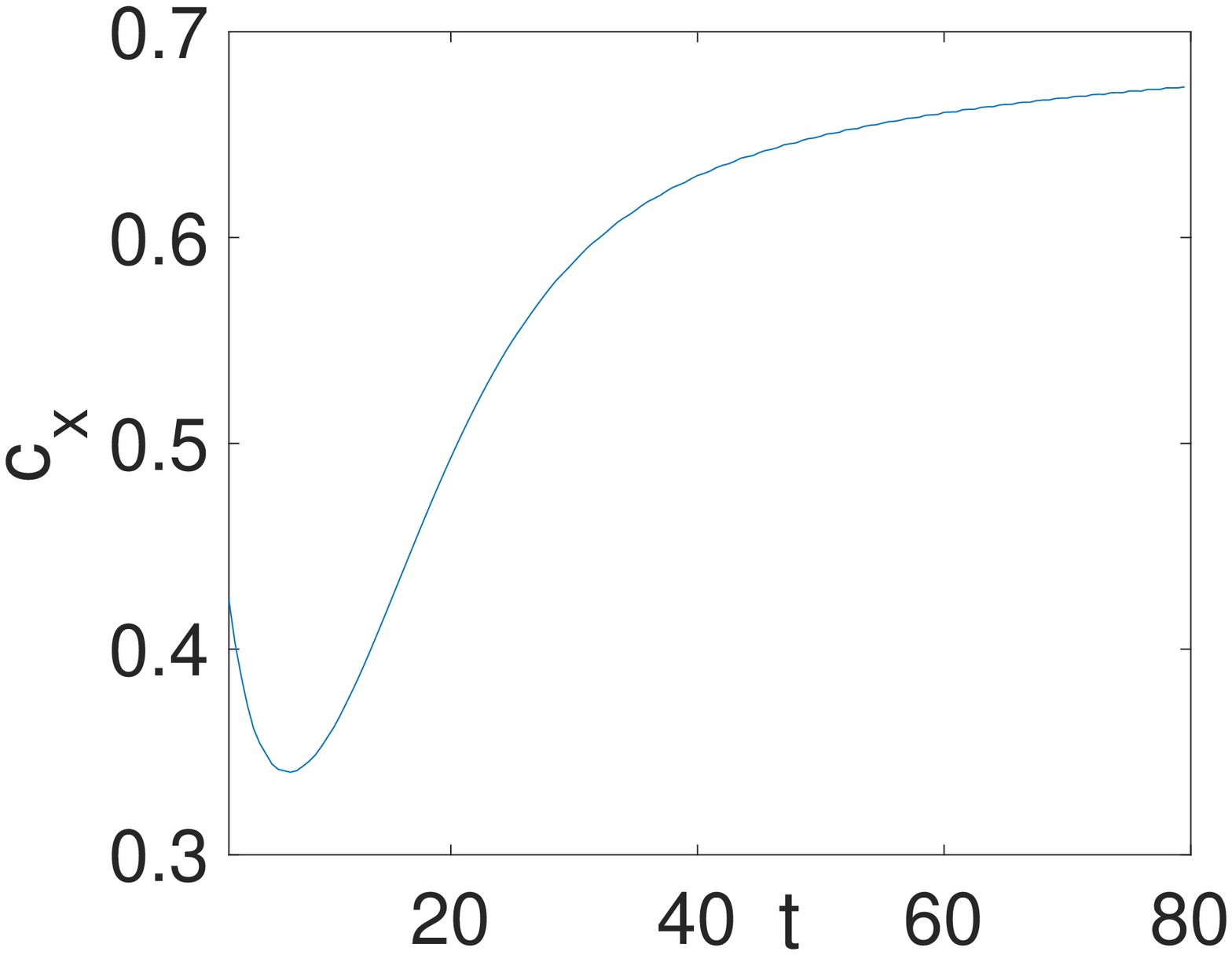}
\caption{The left panel shows the x-position of the center of the front ($x_F$) that starts on the left (and
  moves towards the right) within the
  numerical experiment of a primary tauopathy.
The middle panel shows the corresponding y-position $y_F$
  The speed $c_x$ along the x-axis can
  be
clearly seen in the right panel to approach the asymptotic value of
$c_x=2^{-1/2}$. Correspondingly
the $y$-speed (not shown here) can be seen to approach the limiting
value $c_y=(2 \epsilon)^{-1/2}$.}
\label{evol}
\end{figure}
%%%%%%%%%%%%%%%%%% primary tauopathy using geometry

It can be observed that the scenario described theoretically is
realized
here: the symmetry of the coefficients leads to an equally rapid
propagation
of the two (left and right) blobs in both directions with a speed of $\sqrt{1/2}$.
Indeed, we can observe the damage function evolving accordingly
and symmetrically expanding the disorder across the domain in
Fig.~\ref{dark_sp2}.
More concretely, Fig.~\ref{evol} captures one of these fronts as they
start on the left side of the domain and propagate rightward along
the $x$-direction (left panel), while they also expand along the
$y$-direction (middle panel).
Indeed, here, the simulation involves a factor of $\epsilon=0.01$,
leading to a tenfold reduction of the corresponding speed along the
$y$-direction. It can be seen
that our numerical evaluation of the associated speed, after a
transient
(which can also be observed in the left and middle panels),
settles in the vicinity of its anticipated asymptotic value (right
panel
of Fig.~\ref{evol}). Importantly, also, however, we observe in
Fig.~\ref{dark_sp}
the formation of the co-existing (4th) state of the two toxic proteins
A$\beta$ and $\tau$P as the prevalent state where the two populations
overlap. This can be especially discerned in the bottom left and
bottom
right panels of the figure where the higher concentration of the toxic
$\tau$P clearly illustrates the relevant state (recall that the
A$\beta$
does not modify its equilibrium concentration in the presence of
$\tau$P).
Notice also that the damage function, as defined herein,
also does not appear to feature
an immediately discernible signature of the co-existence state, as per Fig.~\ref{dark_sp2}.

\begin{figure}[ht]
\centering
\includegraphics[scale=0.3]{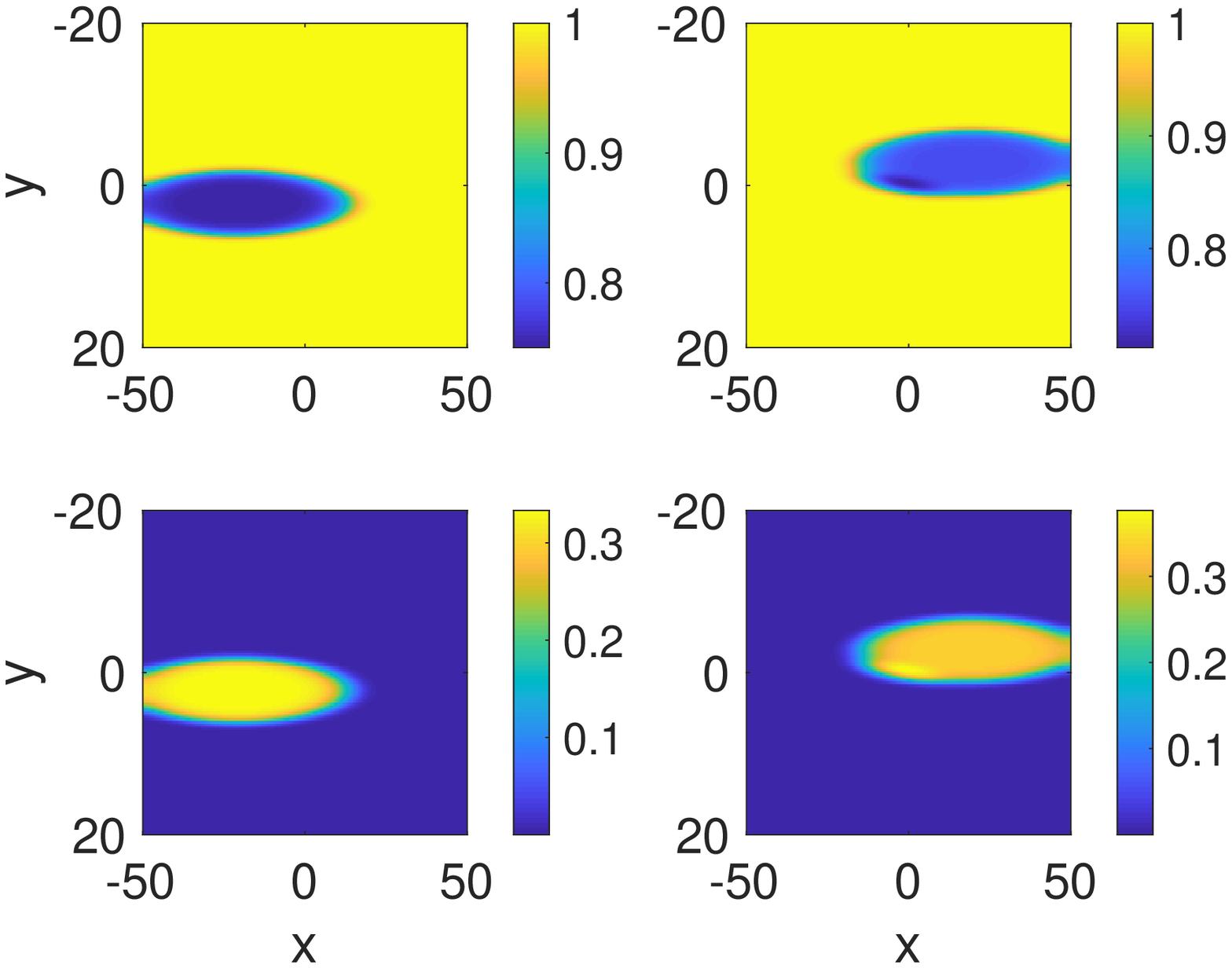}
\includegraphics[scale=0.3]{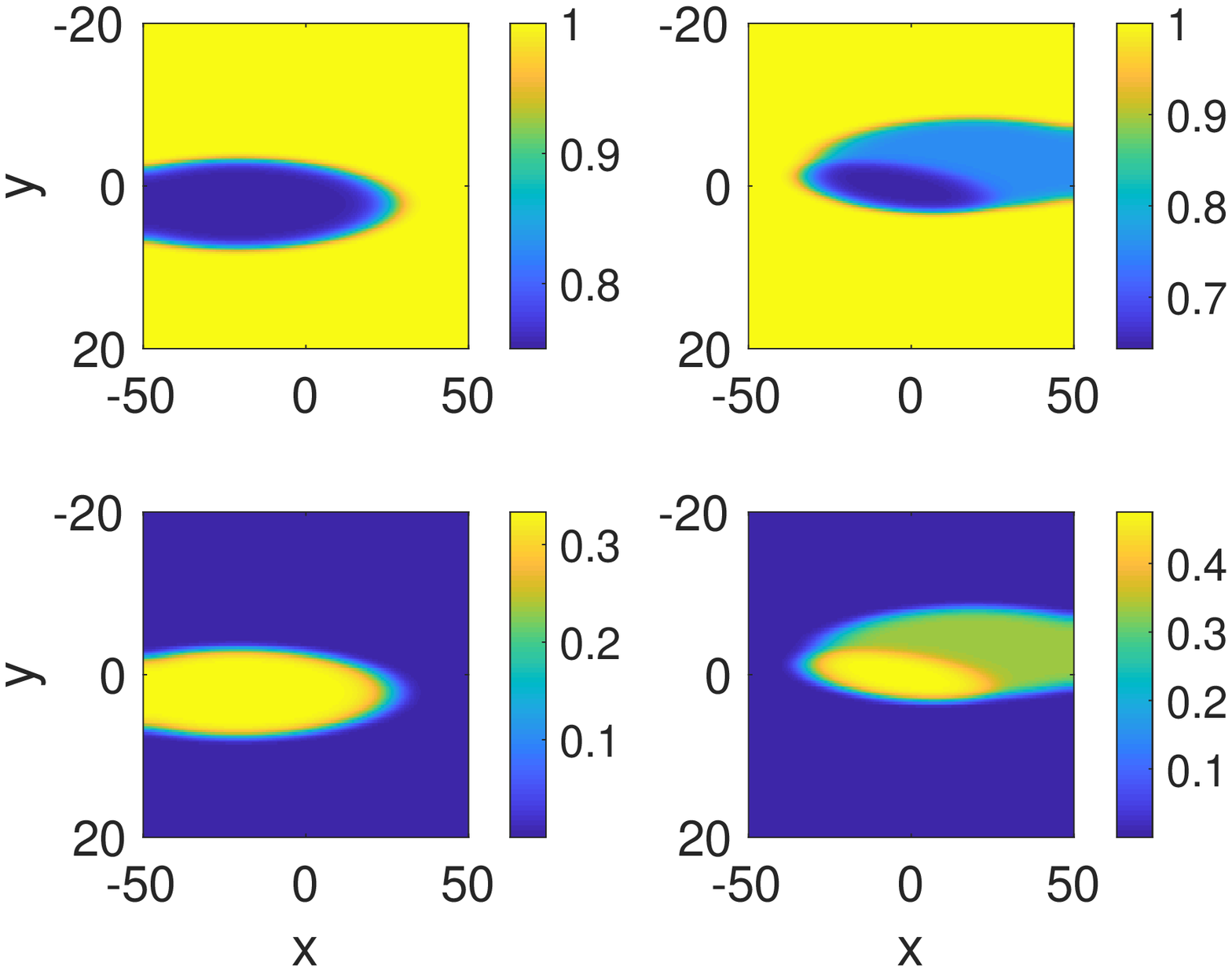}   \\
\includegraphics[scale=0.3]{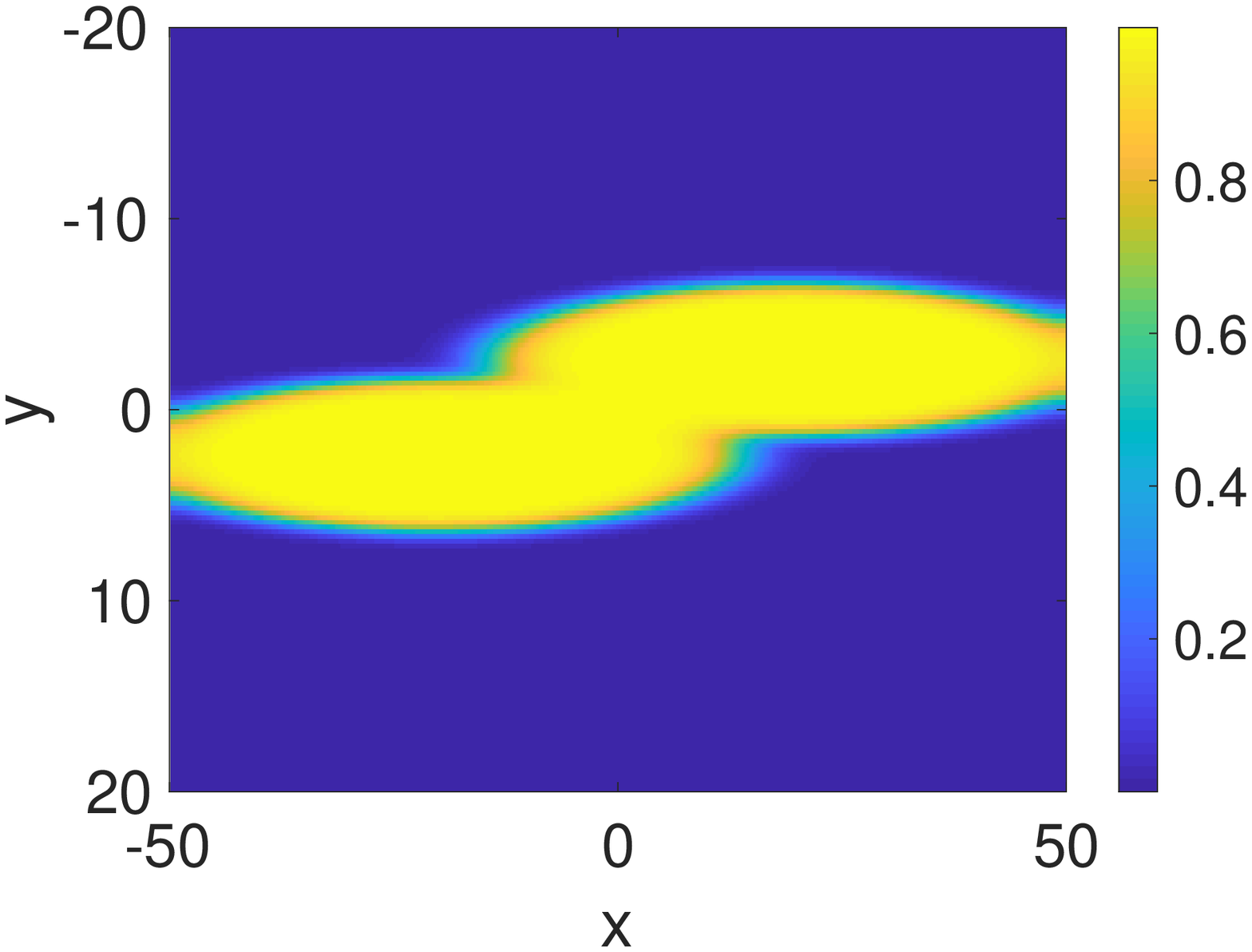}
\includegraphics[scale=0.3]{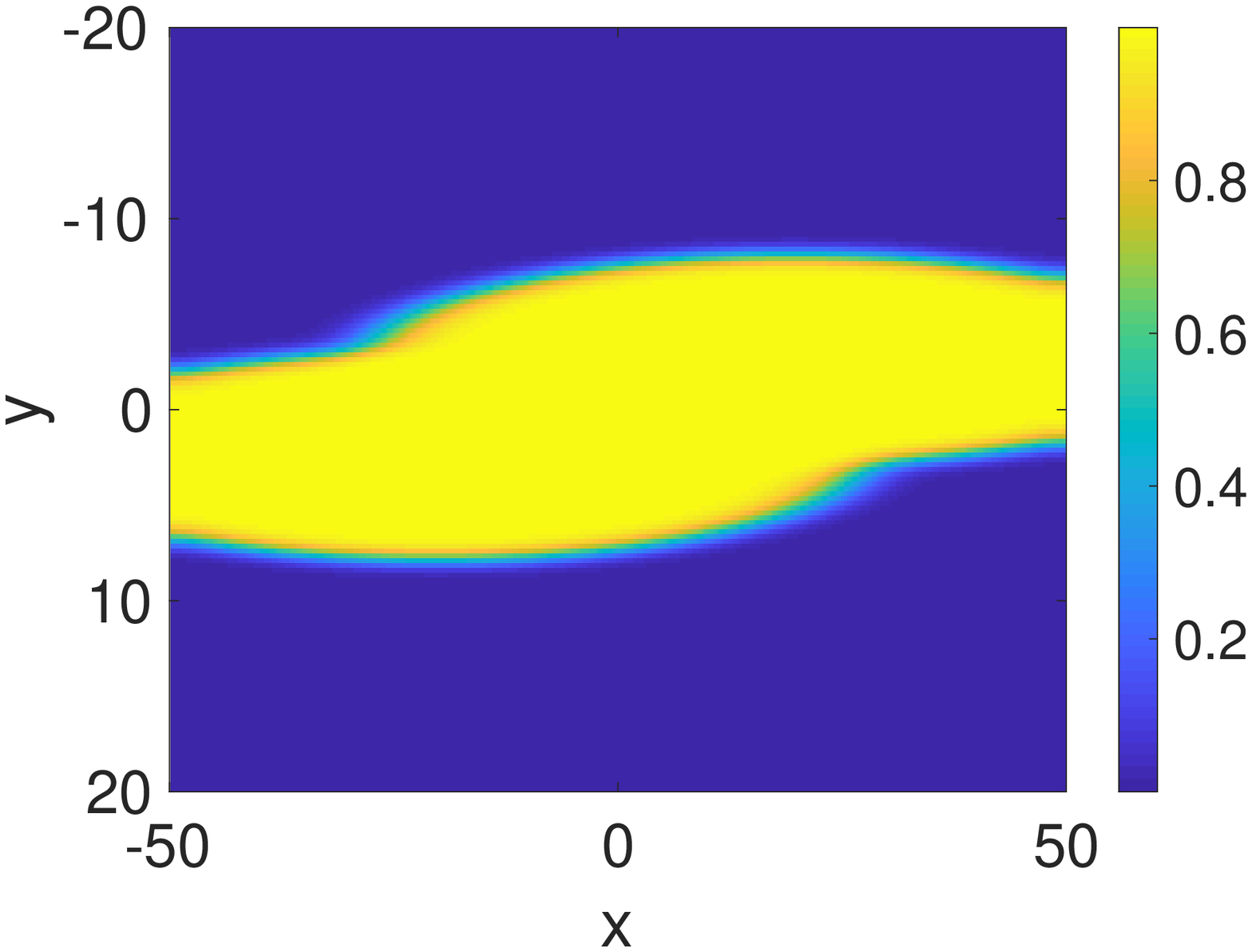}
\caption{Similar to the results of the primary tauopathy case, but now
in a case where the geometry/two-dimensionality of the initial
configuration
matters more: initially the waves of toxicity of A$\beta$ and $\tau$P
are offset as per Eqs.~(\ref{in1b}) and~(\ref{in2b}). The four fields
are shown for
$t=60$ (top left $2 \times 2$ panels) and $t=80$ (top right $2 \times 2$ panels)
along with their corresponding damage variable
spatio-temporal spread (bottom panels).}
\label{dark_sp_1b}
\end{figure}

To explore the effects of geometry and two-dimensionality of the
system,
we now turn to the consideration of a scenario where the initial toxicity of the
A$\beta$ and $\tau$P are not ``aligned''. In this case, while
we retain the initially uniform profile in the healthy populations
of the relevant biomarkers, we offset vertically the corresponding toxic
initial populations as follows:
\begin{eqnarray}
  \tilde{u}(x,y,0)&=&\frac{1}{3} {\rm sech}^2\left((x+20)^2 + 5 (y-2.5)^2\right)
  \label{in1b}
  \\
  \tilde{v}(x,y,0)&=&\frac{1}{3} {\rm sech}^2\left((x-20)^2 + 5 (y+2.5)^2\right)
                      \label{in2b}
\end{eqnarray}
In this case too, during the early stages, the propagation
of
the neurodegenerative waves (the one connecting the 1st and the 2nd
homogeneous state on the left and the one connecting the 1st and the
3rd such on the right) occurs principally along quasi-one-dimensional
``corridors'' within the system. 
As can be seen in Fig.~\ref{dark_sp_1b}, however, at later times, as
these waves spread in the lateral direction, they interact and form
an ``oblique'' front. Here, the co-existent state of toxicity of the
two species dominates, leading to an expansion of the relevant
front in both directions. This oblique interaction pattern also
affects
the spread of the corresponding damage function as can be observed
in the bottom panels of the figure. Once again, the latter bears no
discernible features of the toxic co-existence associated with the
4th equilibrium state (in
comparison
to the 2nd or 3rd one). Still, the expanding front of co-existent
toxicity
is especially evident in the right column of the snapshots shown
(and even more so in the movies of~\cite{dropbox}).

%%%%%% secondary tauopathy

\begin{figure}[htbp]
\centering
\includegraphics[scale=0.3]{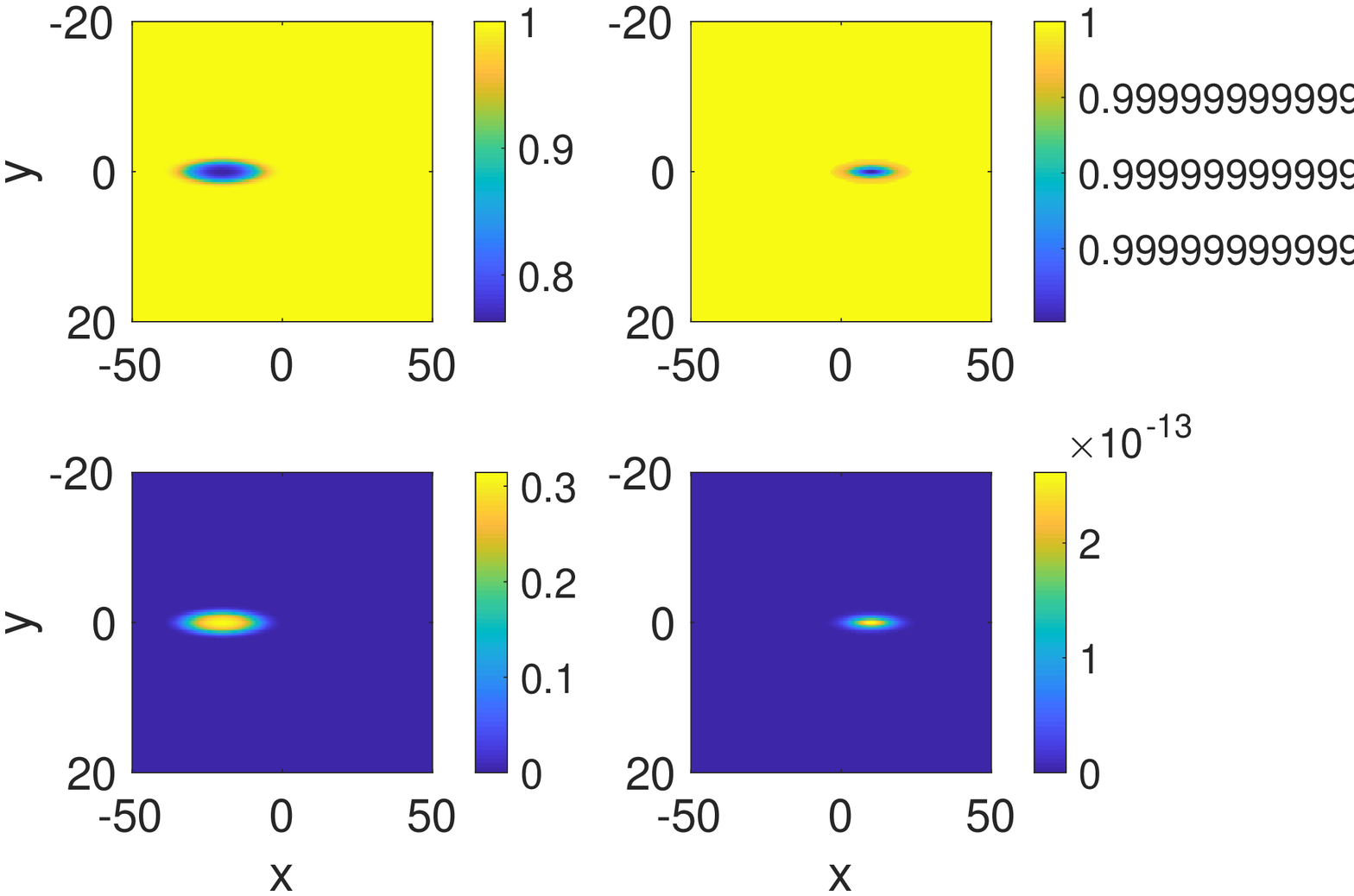}
\includegraphics[scale=0.3]{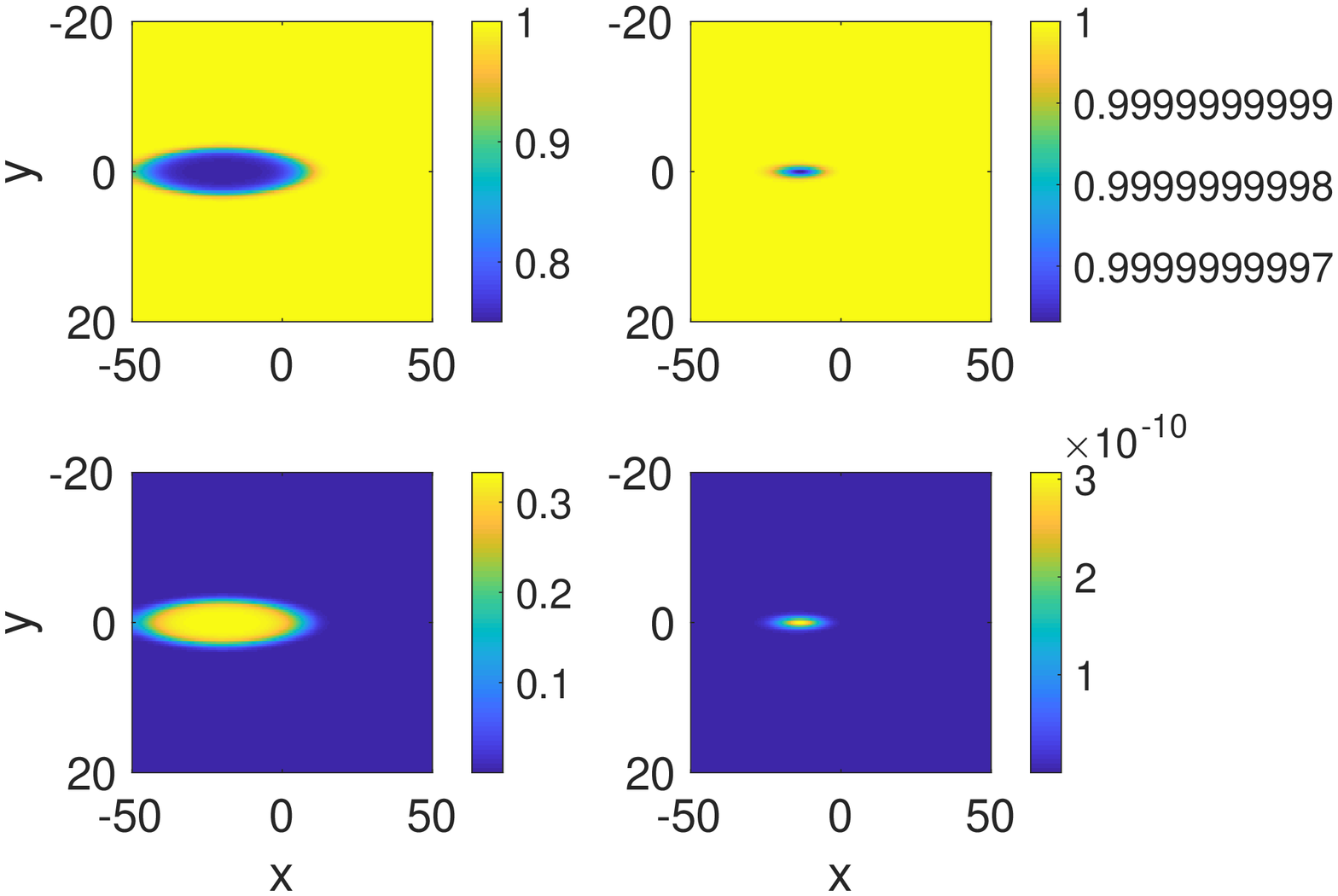}   \\
\includegraphics[scale=0.3]{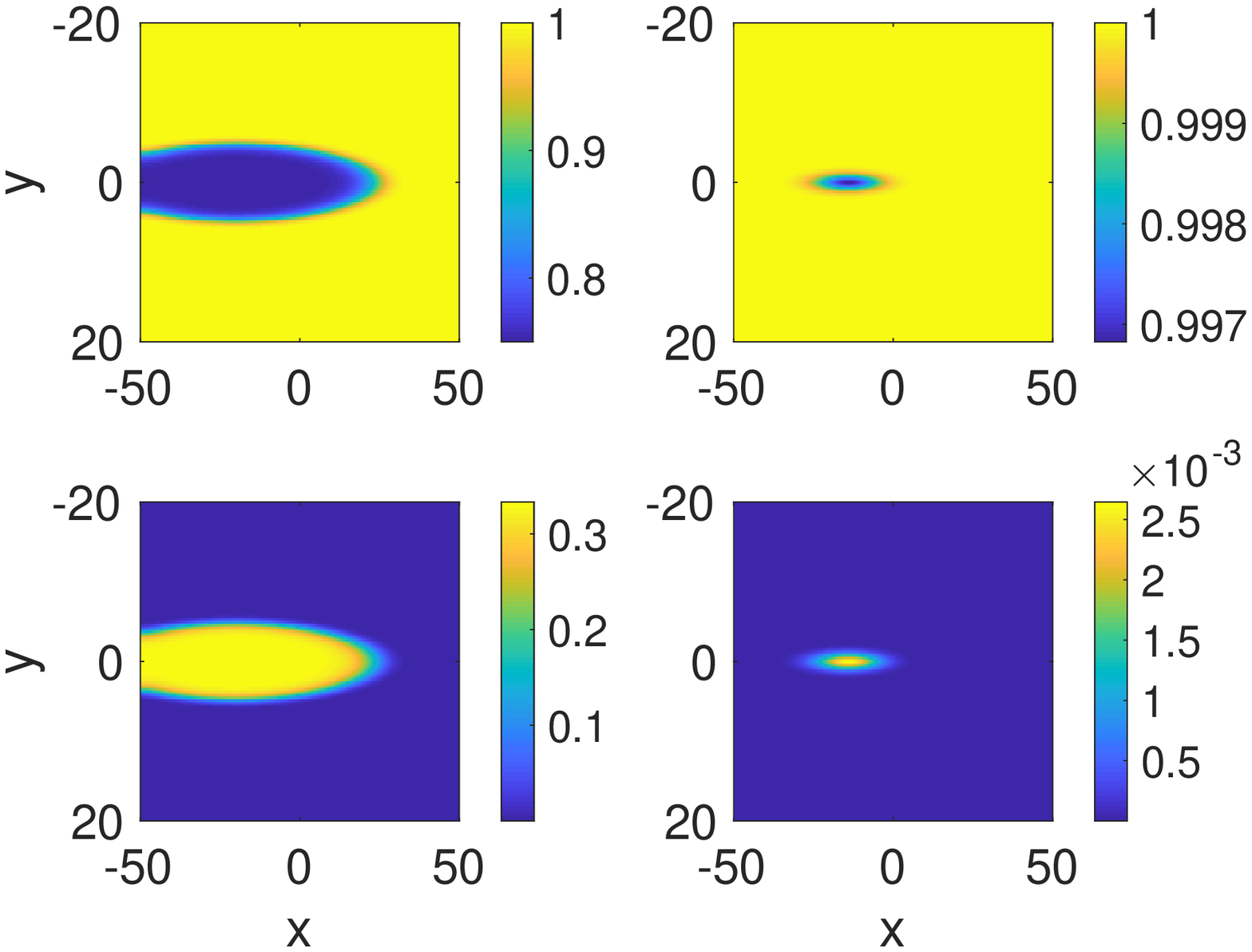}
\includegraphics[scale=0.3]{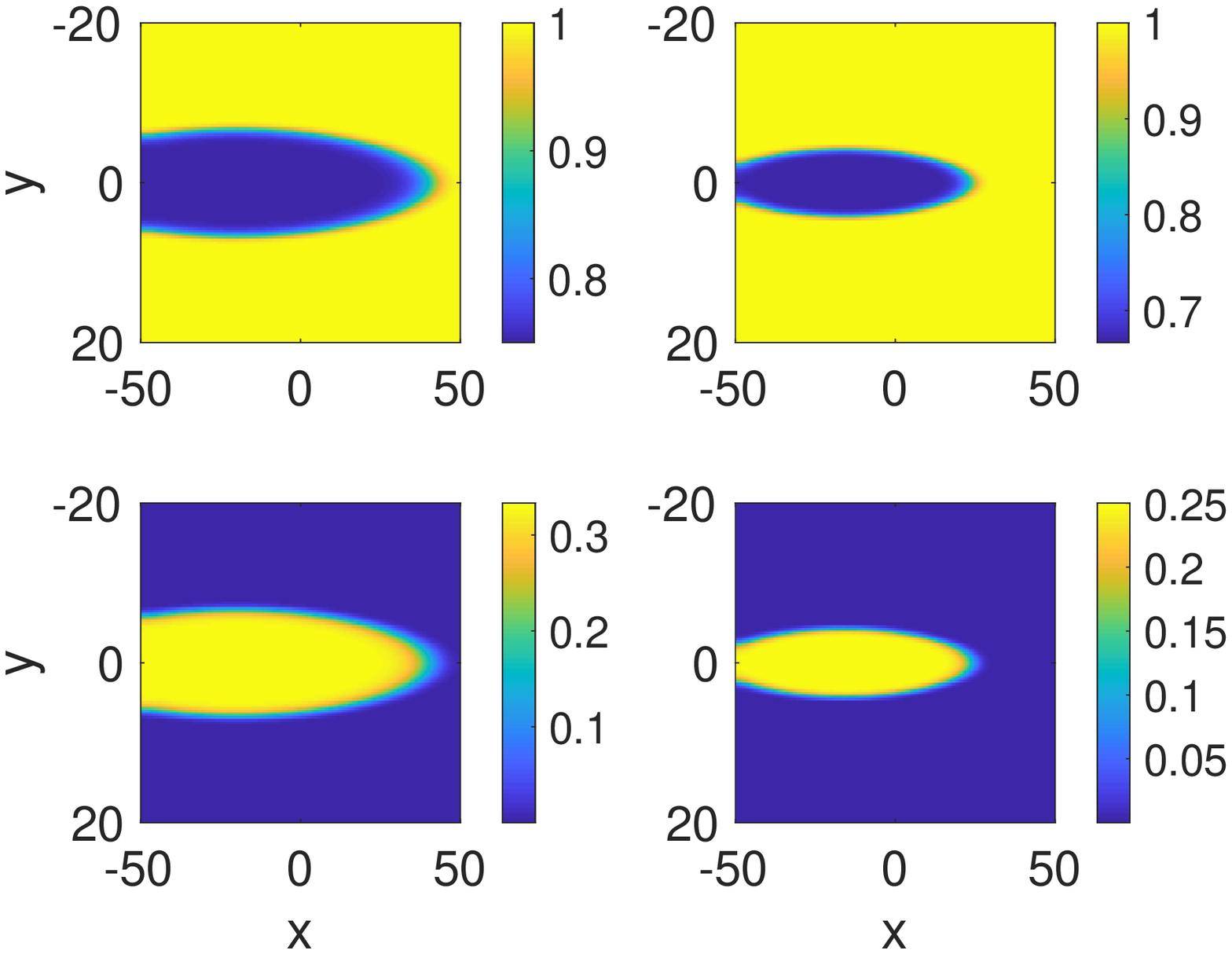}
\caption{Similar to the results of the primary tauopathy case, but now
in the scenario of a secondary tauopathy and for $t=25$, $50$, $75$
and $100$. Notice how the toxic $\tau$P
is absent early on, yet it emerges as a result of its
overlap
with toxic A$\beta$ and subsequently grows in a rapidly expanding front.}
\label{dark_sp_2}
\end{figure}

\begin{figure}[ht]
\centering
\includegraphics[scale=0.25]{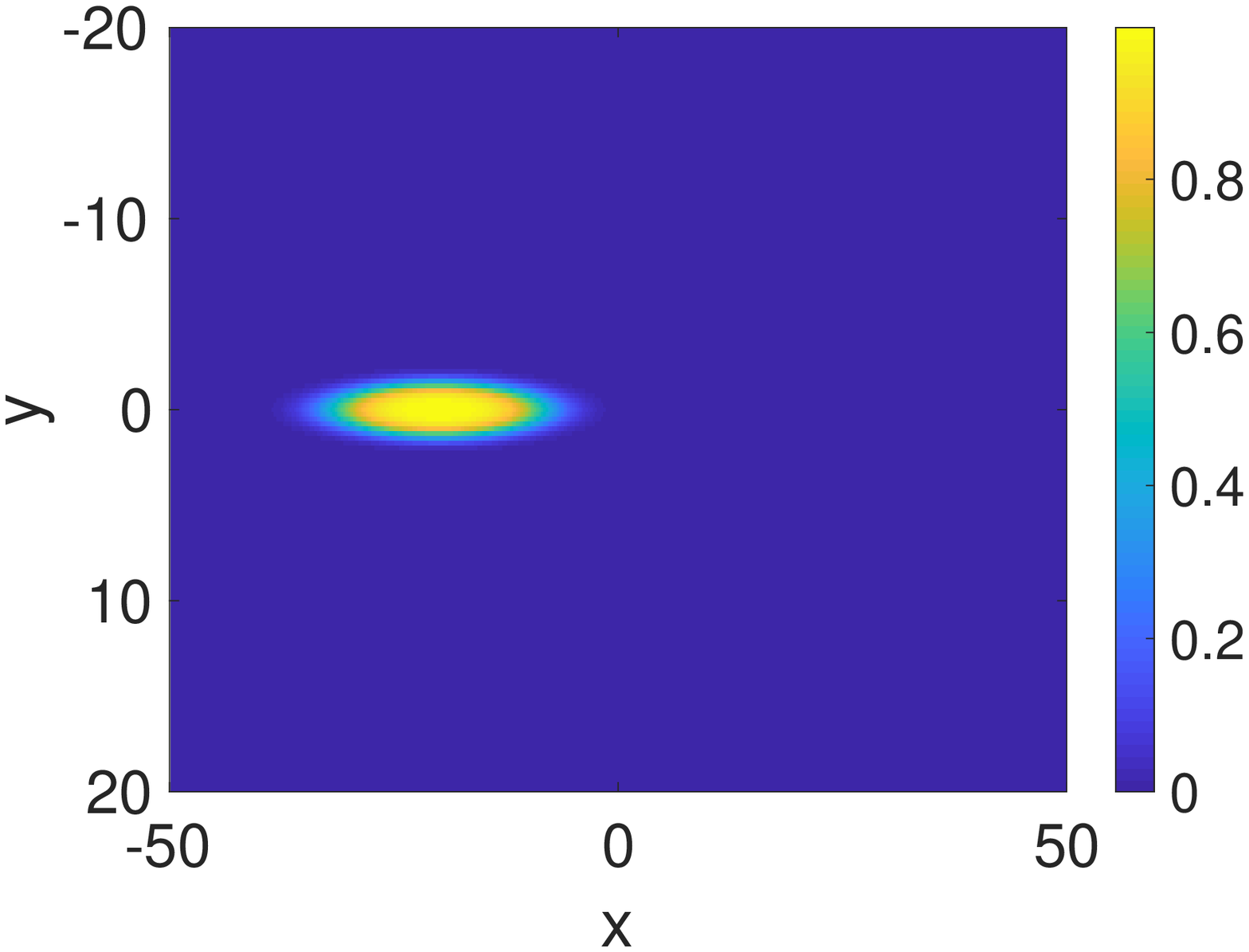}
\includegraphics[scale=0.25]{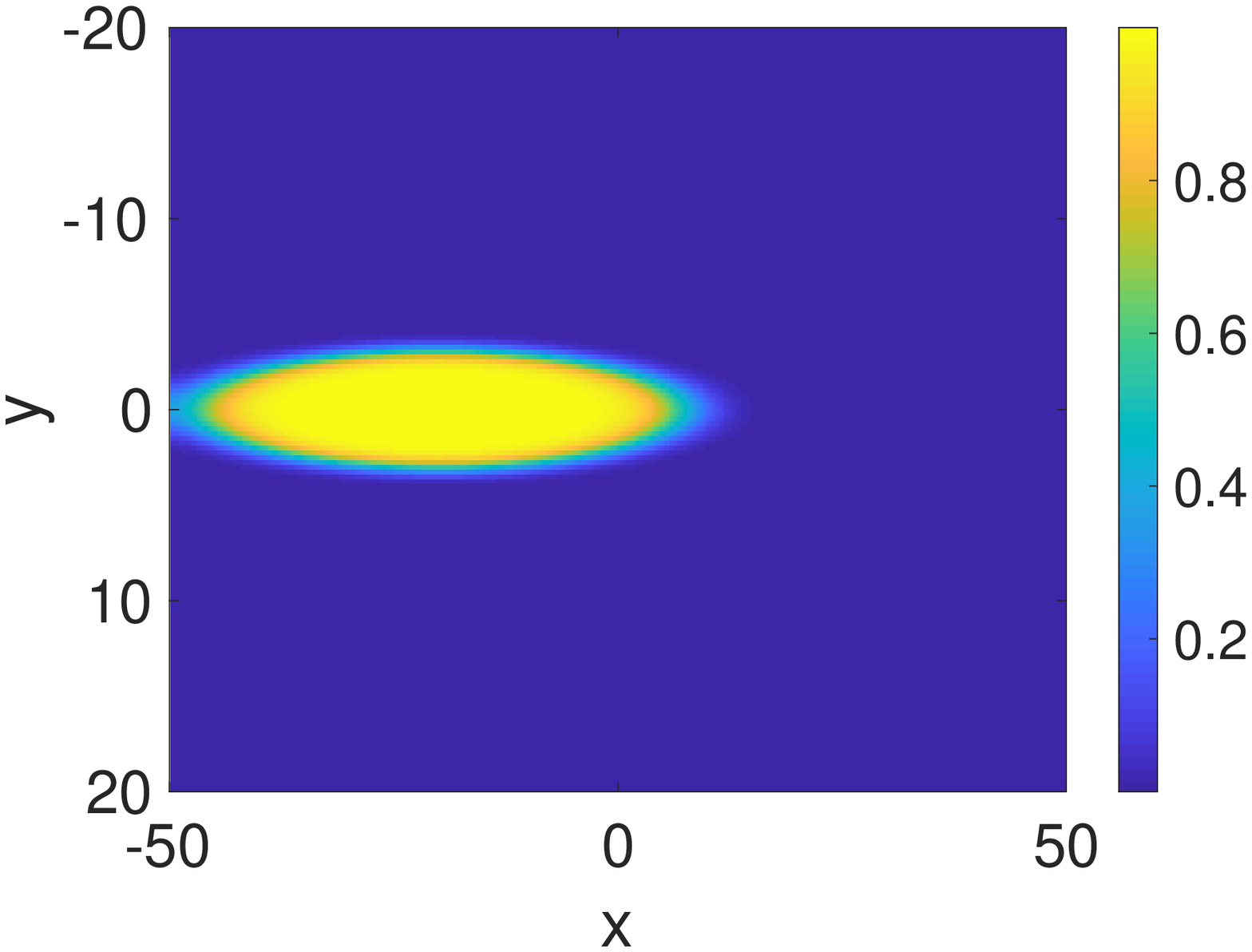}
\includegraphics[scale=0.25]{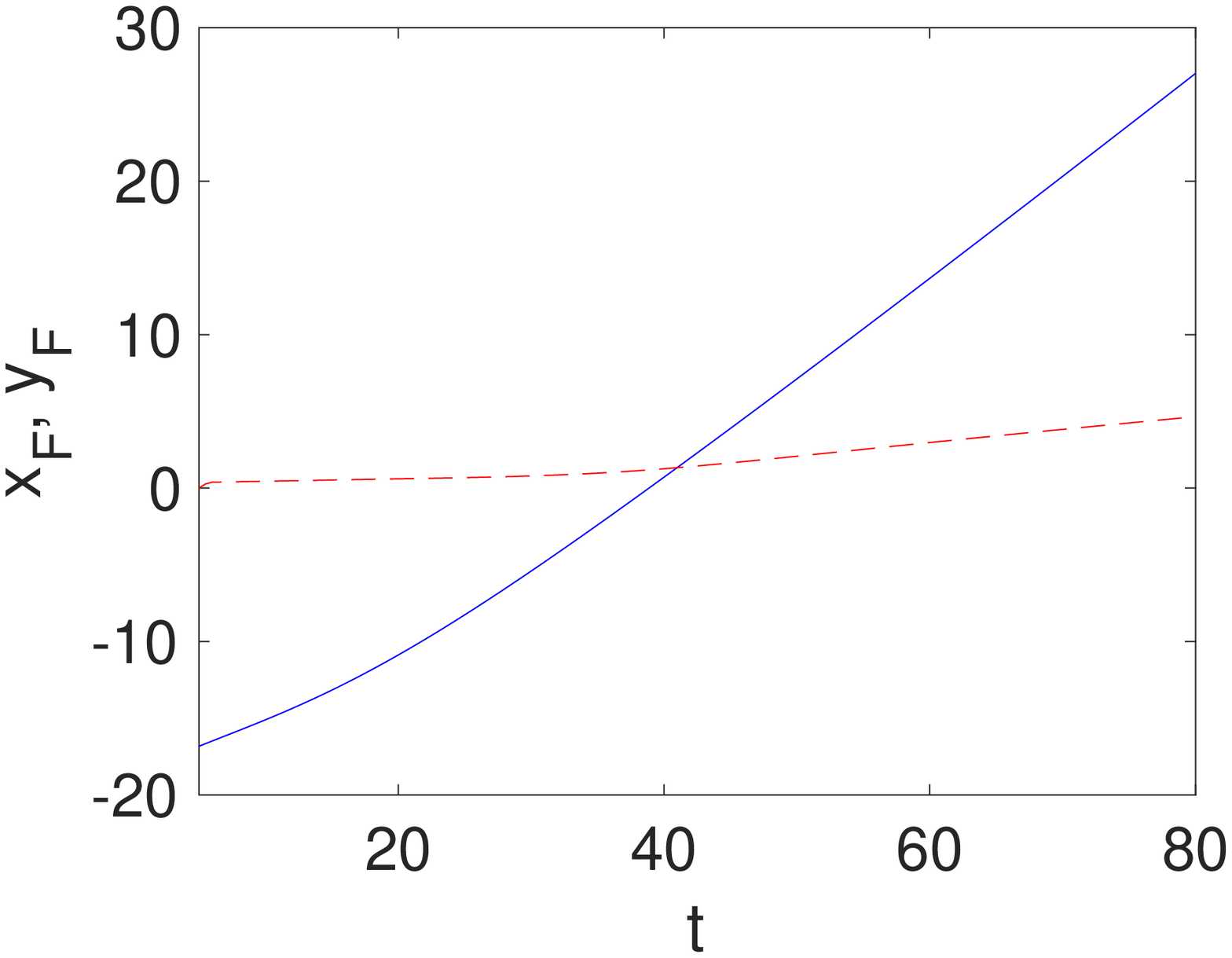} \\
\includegraphics[scale=0.25]{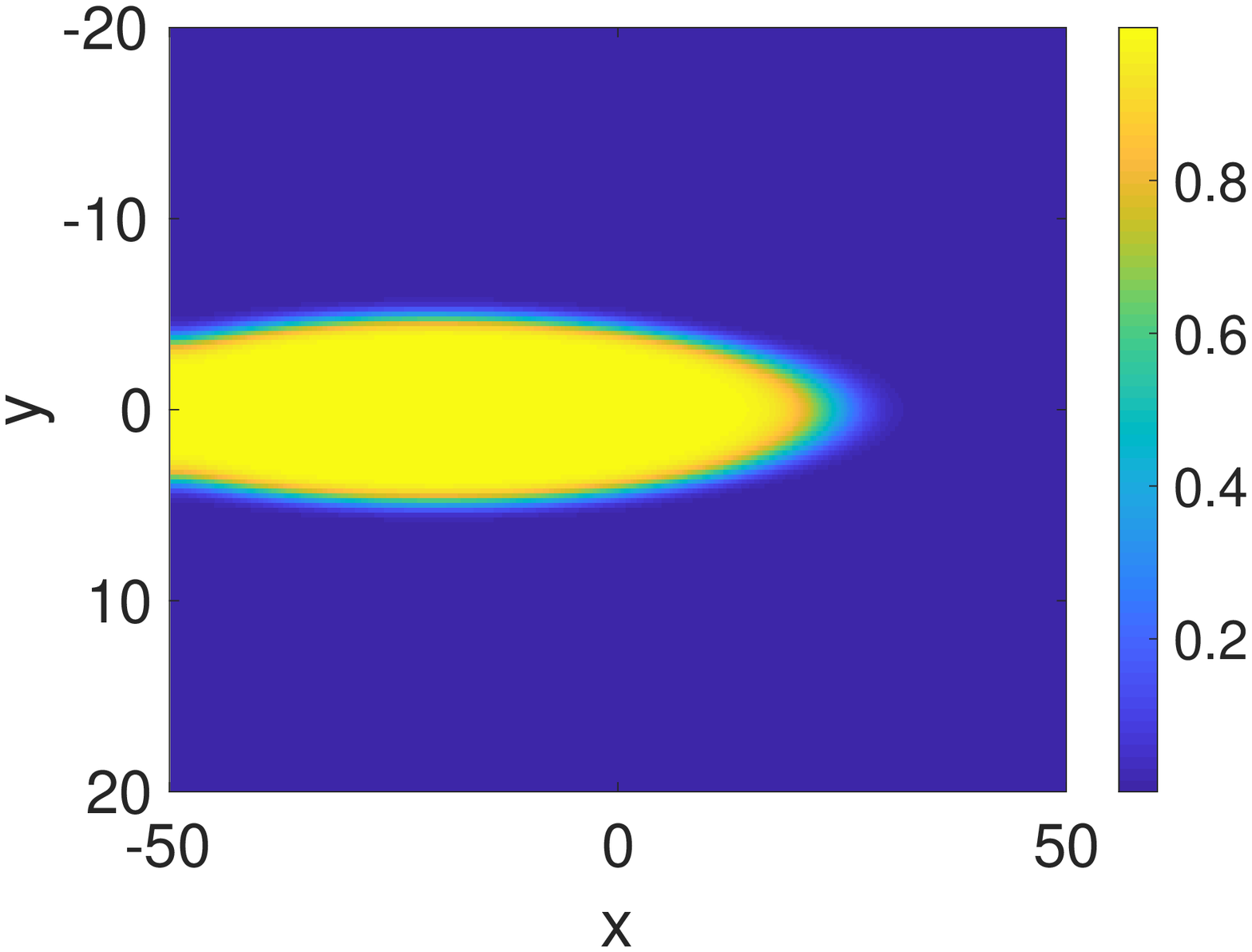}
\includegraphics[scale=0.25]{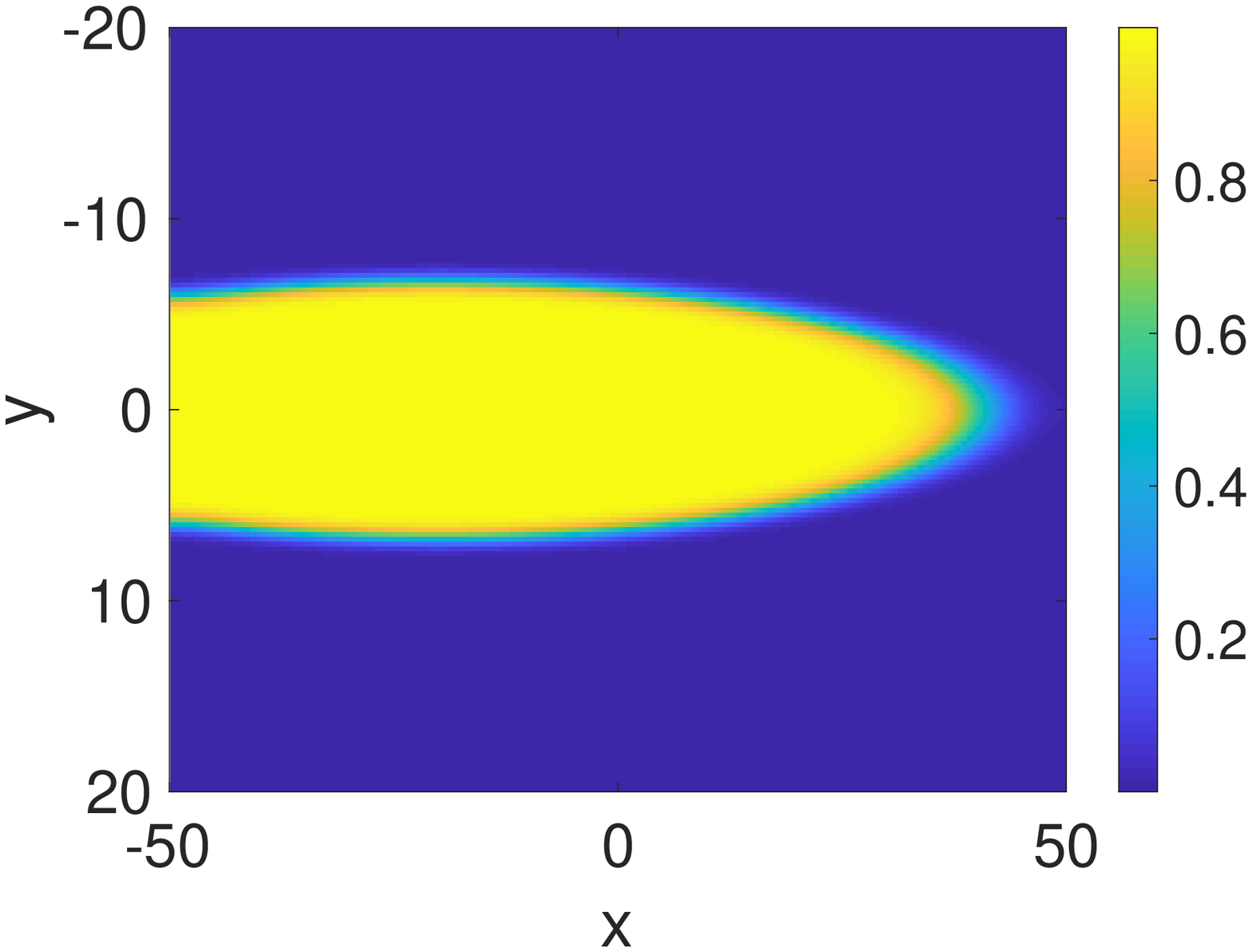}
\includegraphics[scale=0.25]{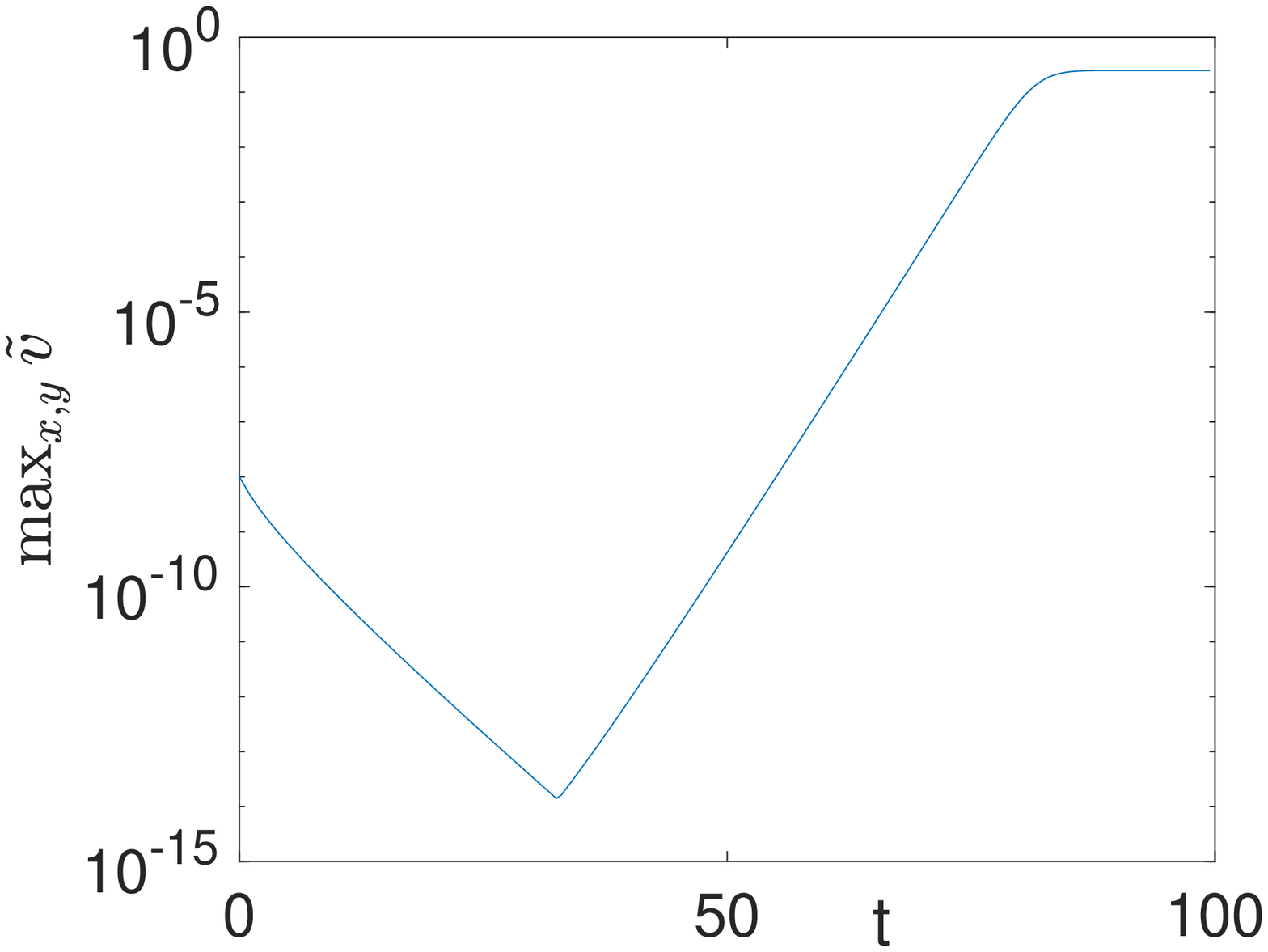}
\caption{The left set of panels involves the damage function at
 the same times as above. The top right panel shows the
  expansion
  of the right-moving toxic front of A$\beta$ in the x (solid) and y
  (dashed) axis via its center position $(x_F,y_F)$.
  The secondary nature of the tauopathy is evident in
  the bottom right showing how the toxic $\tau$P decays until it
  overlaps
  with the rightward propagating toxic A$\beta$ leading to its rapid
  growth and eventual saturation in the co-existing toxic state.
}
\label{dark_sp2_2}
\end{figure}

\subsection{Secondary Tauopathy}
We now turn to a scenario of secondary tauopathy for which the presence of toxic A$\beta$ is required for toxic $\tau$P.
As discussed in the theory, we select a sufficiently large value of
$b_3=3$, and keep all other coefficients the same except for
$\tilde{b}_1=4/3$, so that the third equilibrium (of solely toxic
$\tau$P) is absent.
In this case, in terms of initial conditions,
the first three components are similar to our original
numerical experiment involving uniform populations for the healthy
biomarkers
and a toxic A$\beta$ population given by Eq.~(\ref{in1}).
However, here the toxic component of the $\tau$P is given by:
\begin{eqnarray}
%  \tilde{u}(x,y,0)&=& 10^{-8}{\rm sech}^2\left((x-10)^2 + 10 y^2\right)
 % \label{in1}
 % \\
  \tilde{v}(x,y,0)&=& 10^{-8} {\rm sech}^2\left((x-20)^2 + 10 y^2\right).
                      \label{in4}
\end{eqnarray}
In this case, a fundamentally different dynamical evolution of the
disorder can be observed. Indeed, the initial stages of the simulation
illustrate a decrease of the toxic levels of $\tau$P (cf. the early
times in Fig.~\ref{dark_sp_2} and also the bottom right panel
of Fig.~\ref{dark_sp2_2},
reporting the maximal concentration thereof). However, over time,
the expansion of the front involving the toxic A$\beta$ eventually
leads to an overlap with the toxic $\tau$P that, in turn, ignites
the nucleation and expansion of the 4th homogeneous state,
the one of co-existent toxicity of the two proteins. The relevant
``droplet'' (of $\tau$P) can be seen to rapidly expand and eventually
catch up to the front of expanding toxic A$\beta$; see the left
and right panels of Fig.~\ref{dark_sp_2}. While this evolution is not
immediately evident in the damage spatio-temporal evolution panels
of Fig.~\ref{dark_sp2_2}, it is clear in the growth and eventual
saturation
of the toxic $\tau$P maximal concentration (bottom right panel
of Fig.~\ref{dark_sp2_2}), as well as in the movies of~\cite{dropbox}.
Notice that we also considered scenarios of non-collinear
propagation in this secondary tauopathy as well (not shown here).
The main difference there was that the non-collinear propagation
delayed the occurrence of overlap between the very weak toxic
$\tau$P pulse and the propagating toxic A$\beta$ front, thus
considerably delaying the emergence and expansion of the 4th
homogeneous state of co-existing toxicity.

%\begin{figure}[htbp]
%\centering
%\includegraphics[scale=0.25]{tau2_xyp_ds.eps}
%%\includegraphics[scale=0.25]{tau2_xyp_t_ds.eps}
%\includegraphics[scale=0.25]{tau2_maxv_ds.eps}
%\caption{}
%\label{evol_2}
%\end{figure}

%\section{FKPP Models} \label{bdg}

%\begin{figure}[htbp]
%\centering
%\includegraphics[scale=0.25]{}
%\caption{}
%\label{dark_sols}
%\end{figure}

%\subsection{Primary Tauopathy Spreading Along a Line}

\begin{figure}[htbp]
\centering
\includegraphics[scale=0.3]{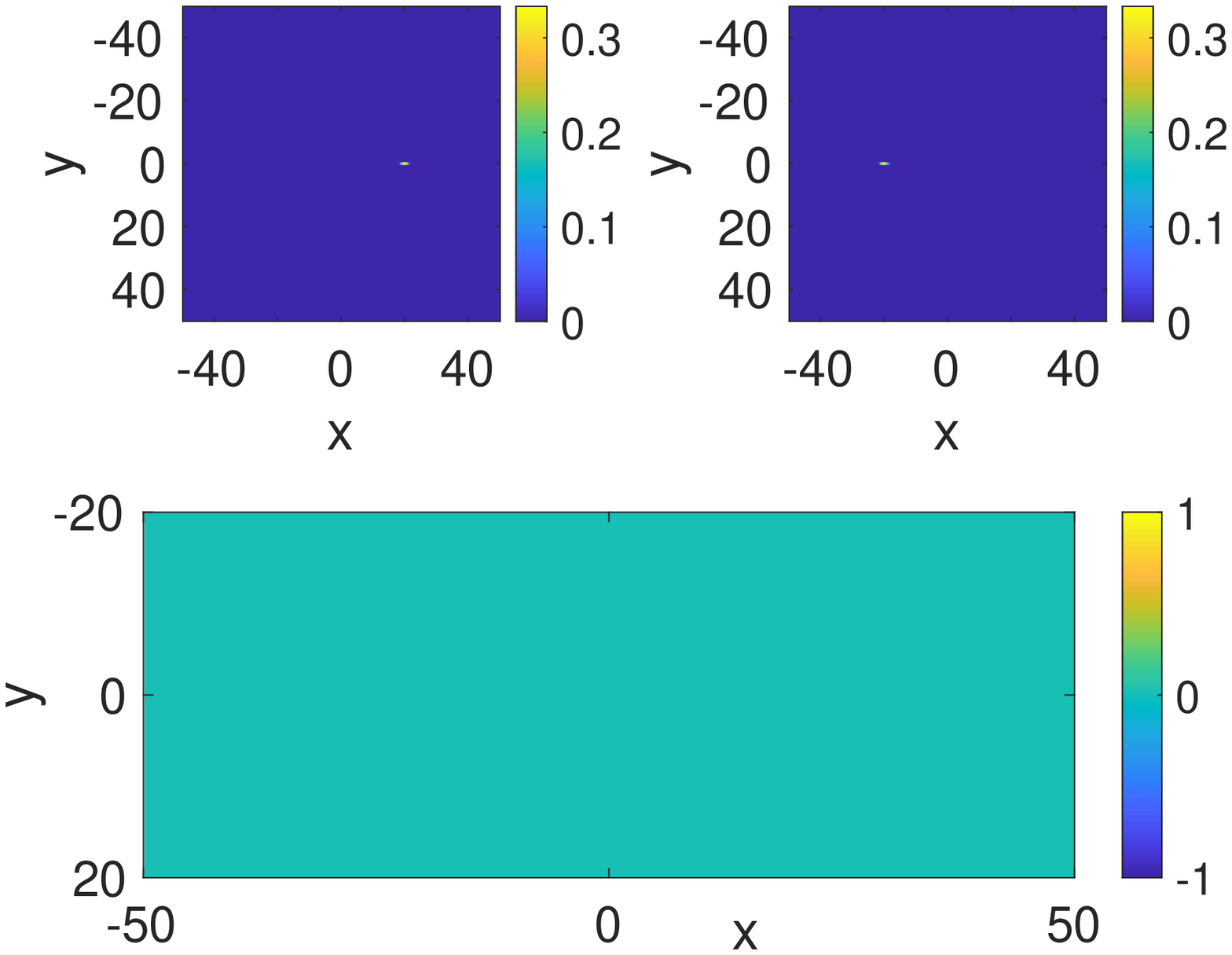}
\includegraphics[scale=0.3]{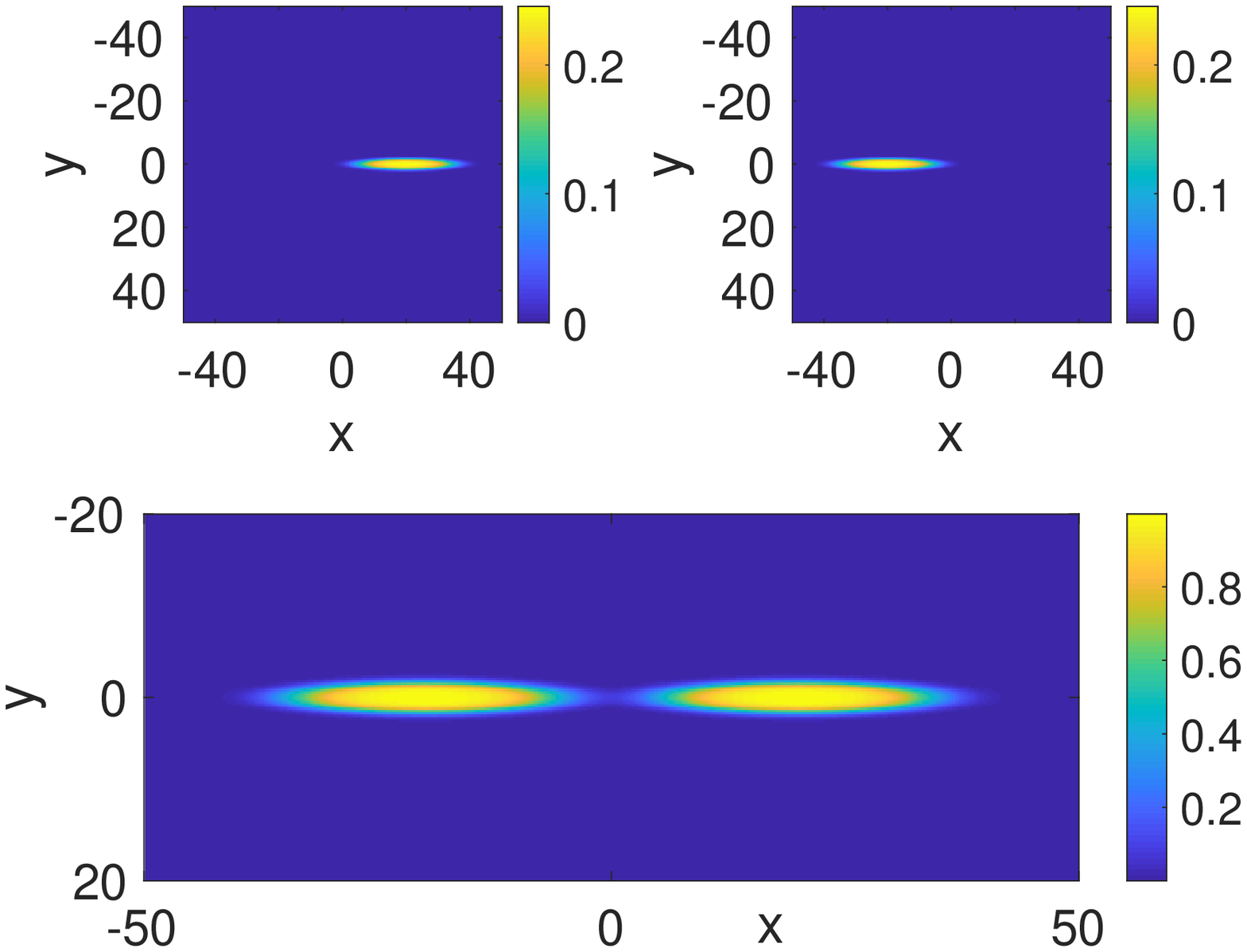}   \\
\includegraphics[scale=0.3]{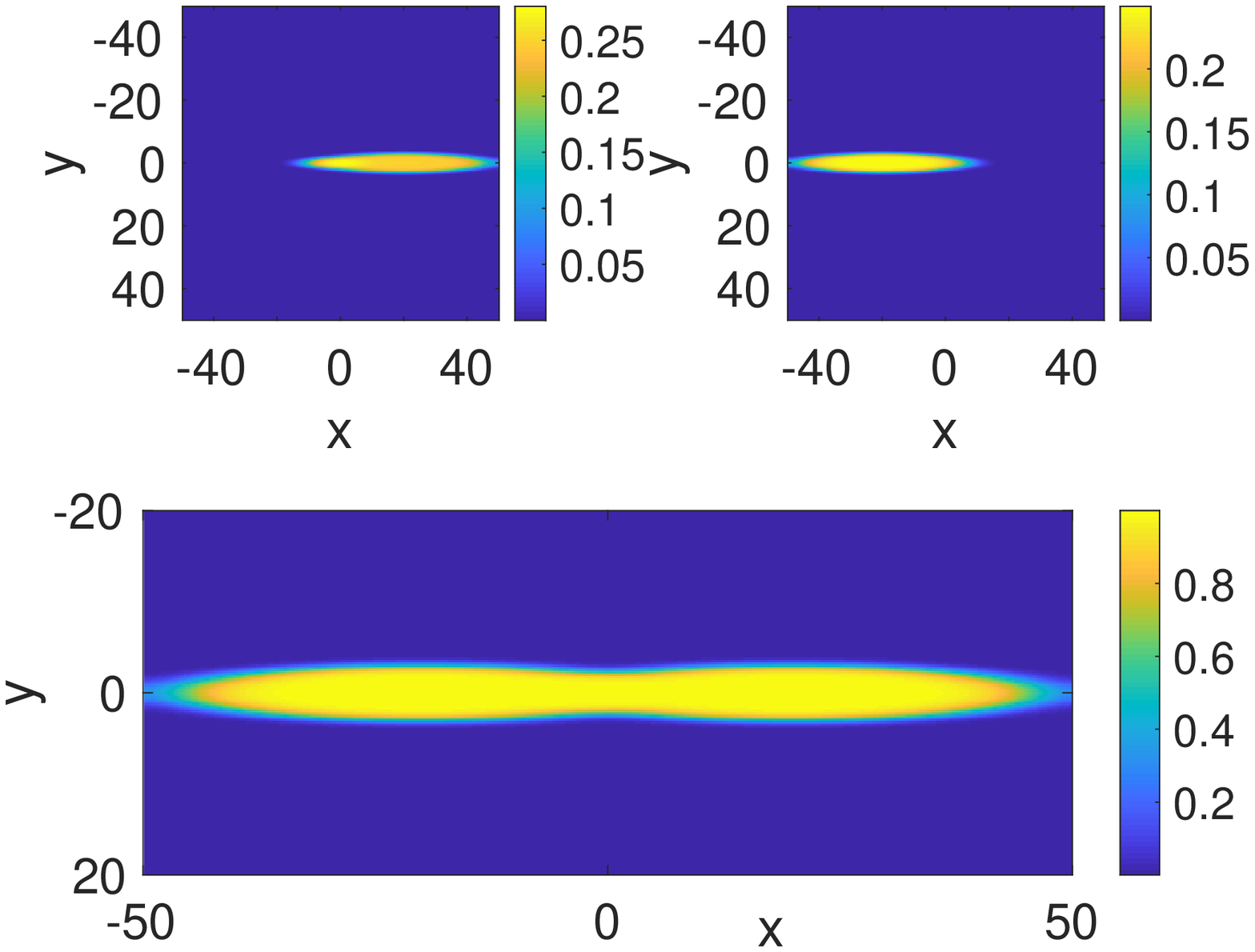}
\includegraphics[scale=0.3]{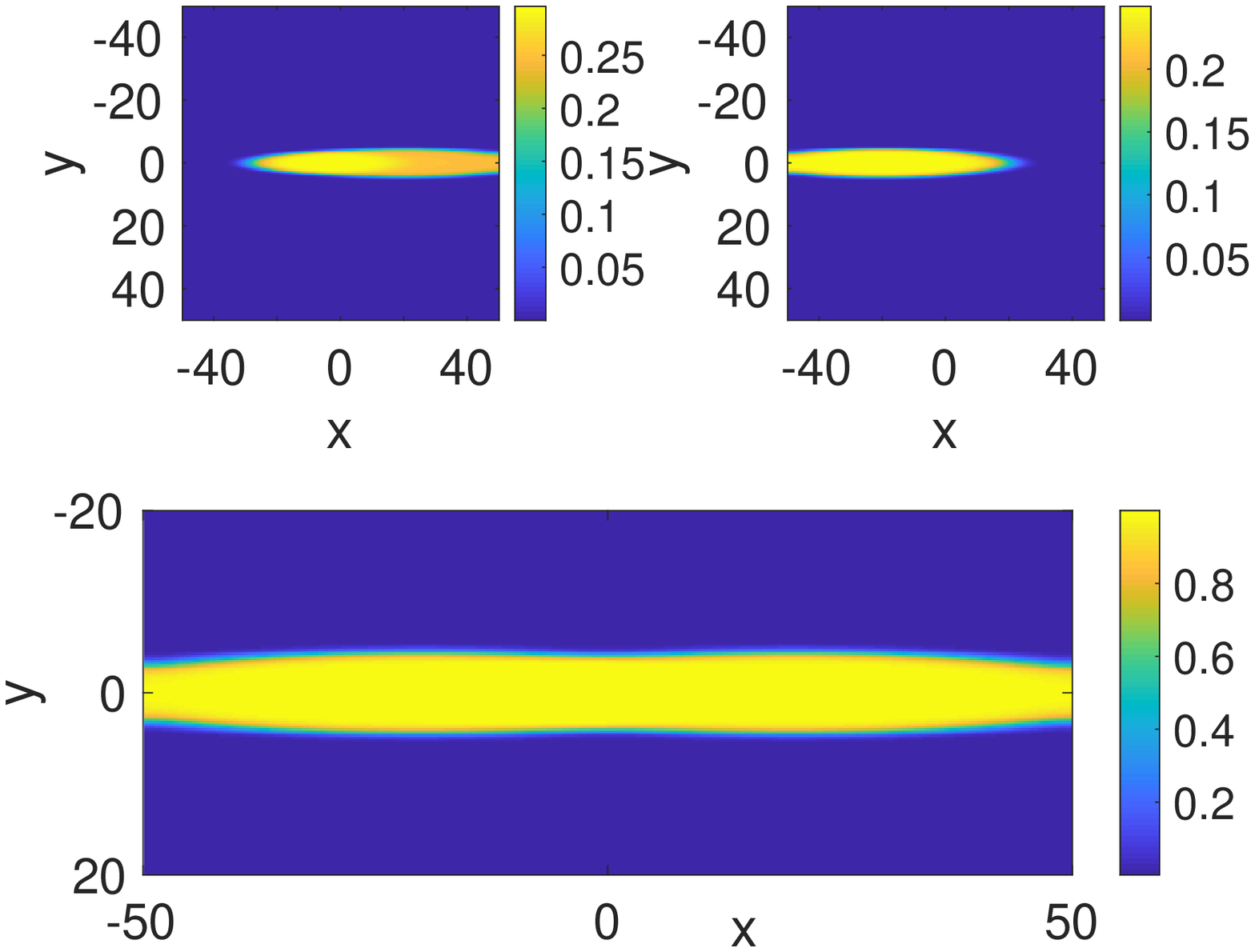}
\caption{Similar to the results of the primary tauopathy case
  (Figs.~\ref{dark_sp} and~\ref{dark_sp2}), but now
  in the scenario of collinear propagation
  within a two-species FKPP-type model only for the toxic components
  of A$\beta$ and $\tau$P. Each triplet shows the two toxic components,
  as well as the damage variable $q$ at the same times as before.}
\label{dark_2c_sp_1}
\end{figure}

\begin{figure}[htbp]
\centering
\includegraphics[scale=0.3]{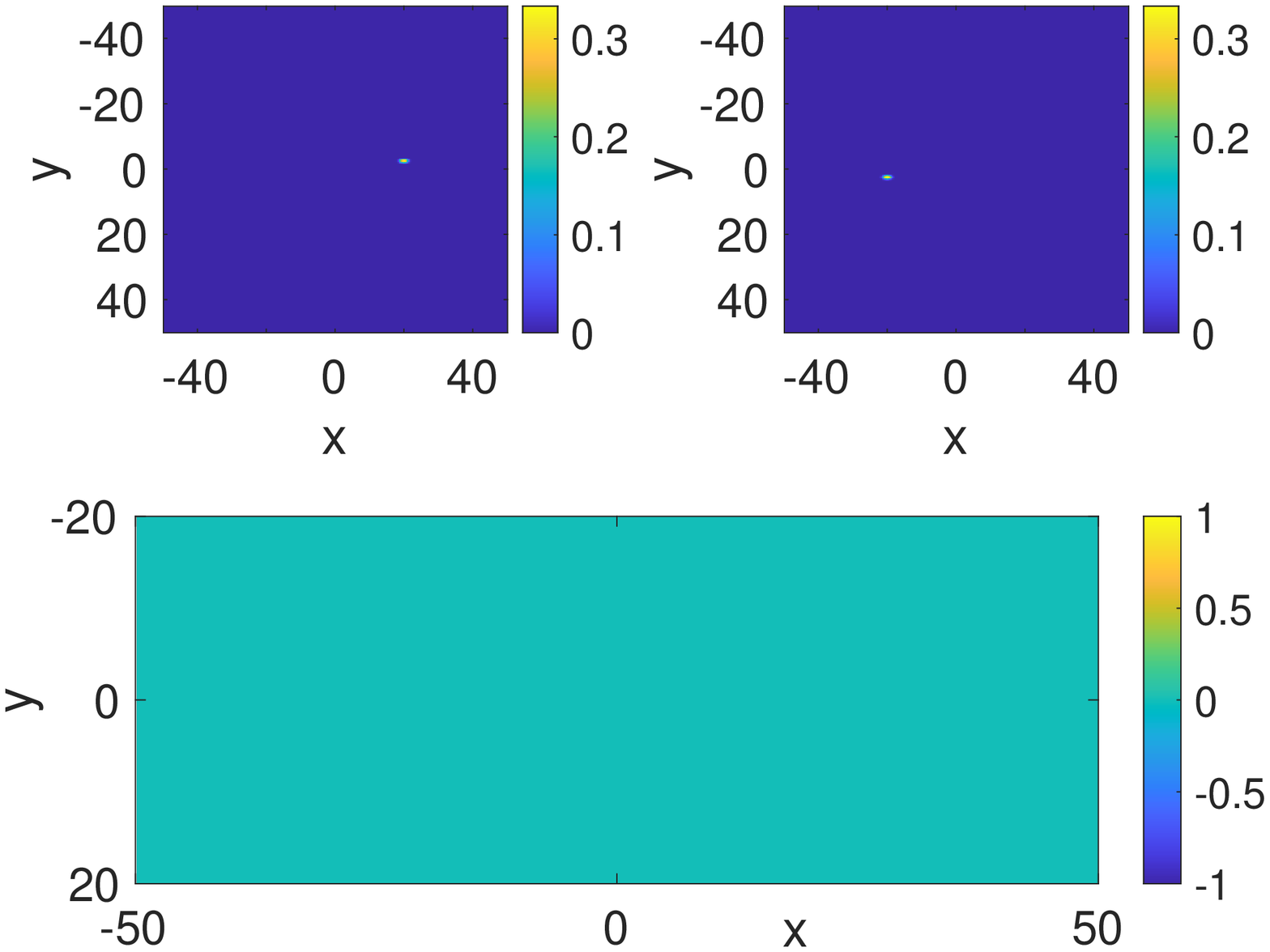}
\includegraphics[scale=0.3]{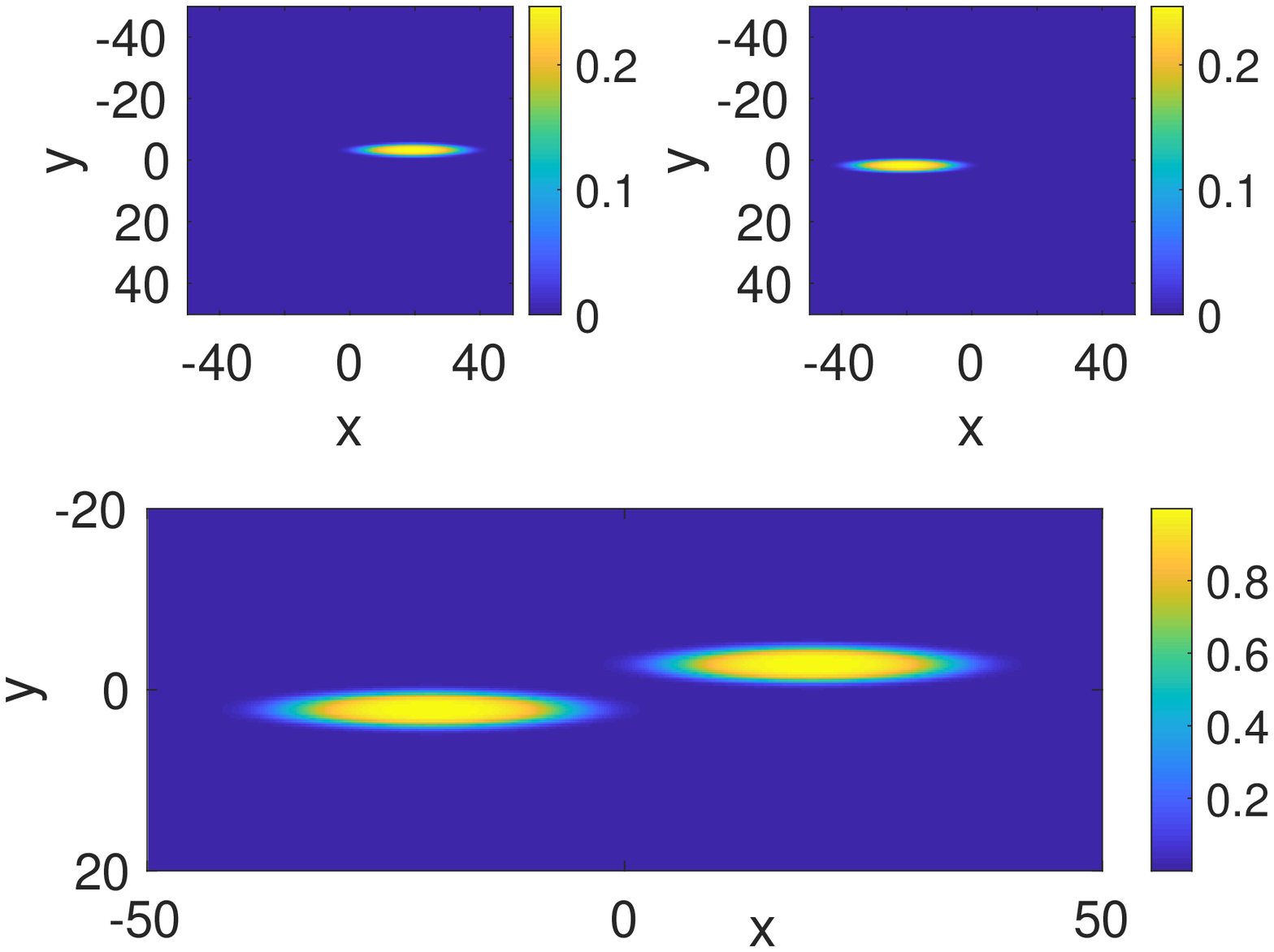}   \\
\includegraphics[scale=0.3]{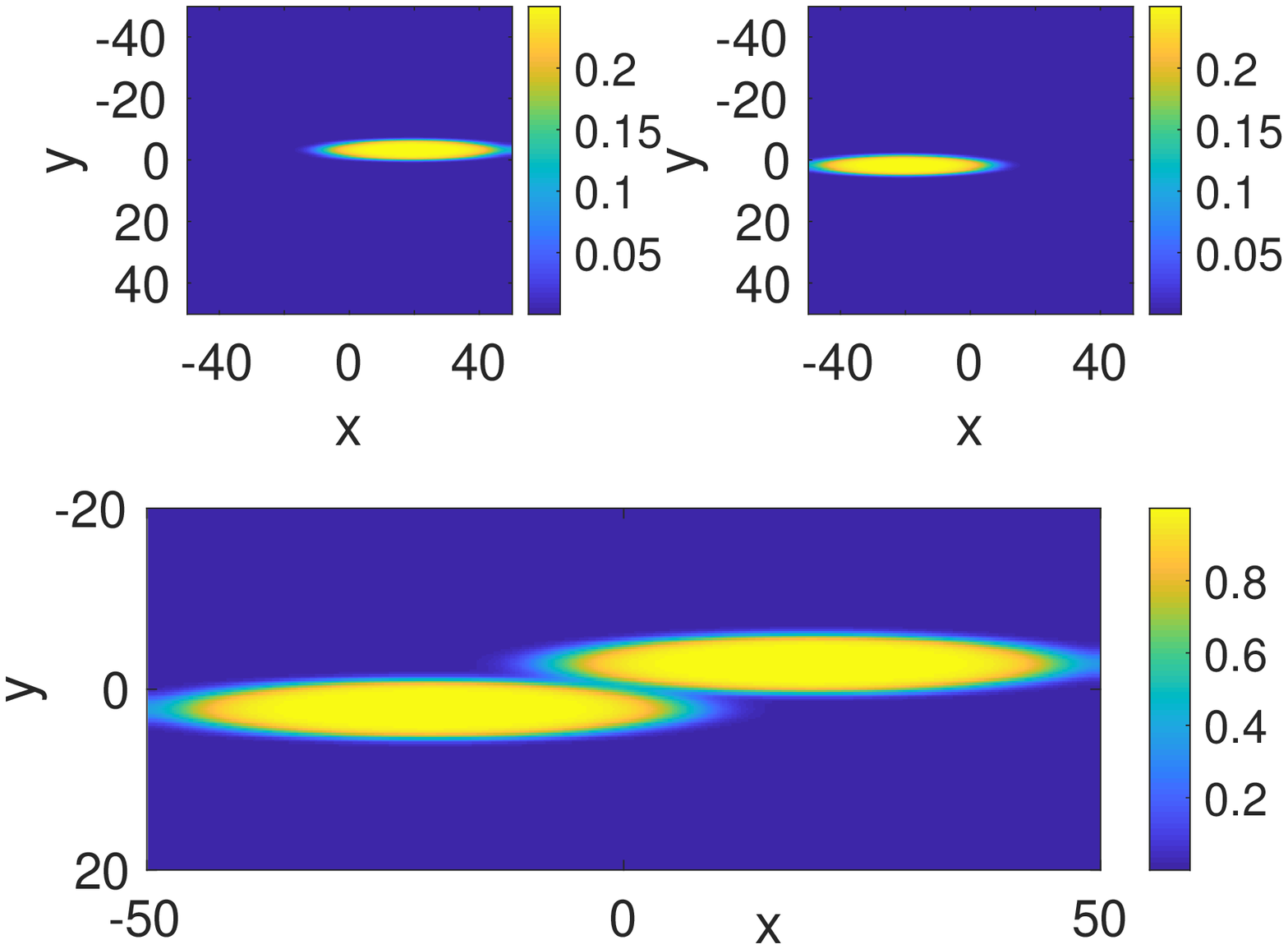}
\includegraphics[scale=0.3]{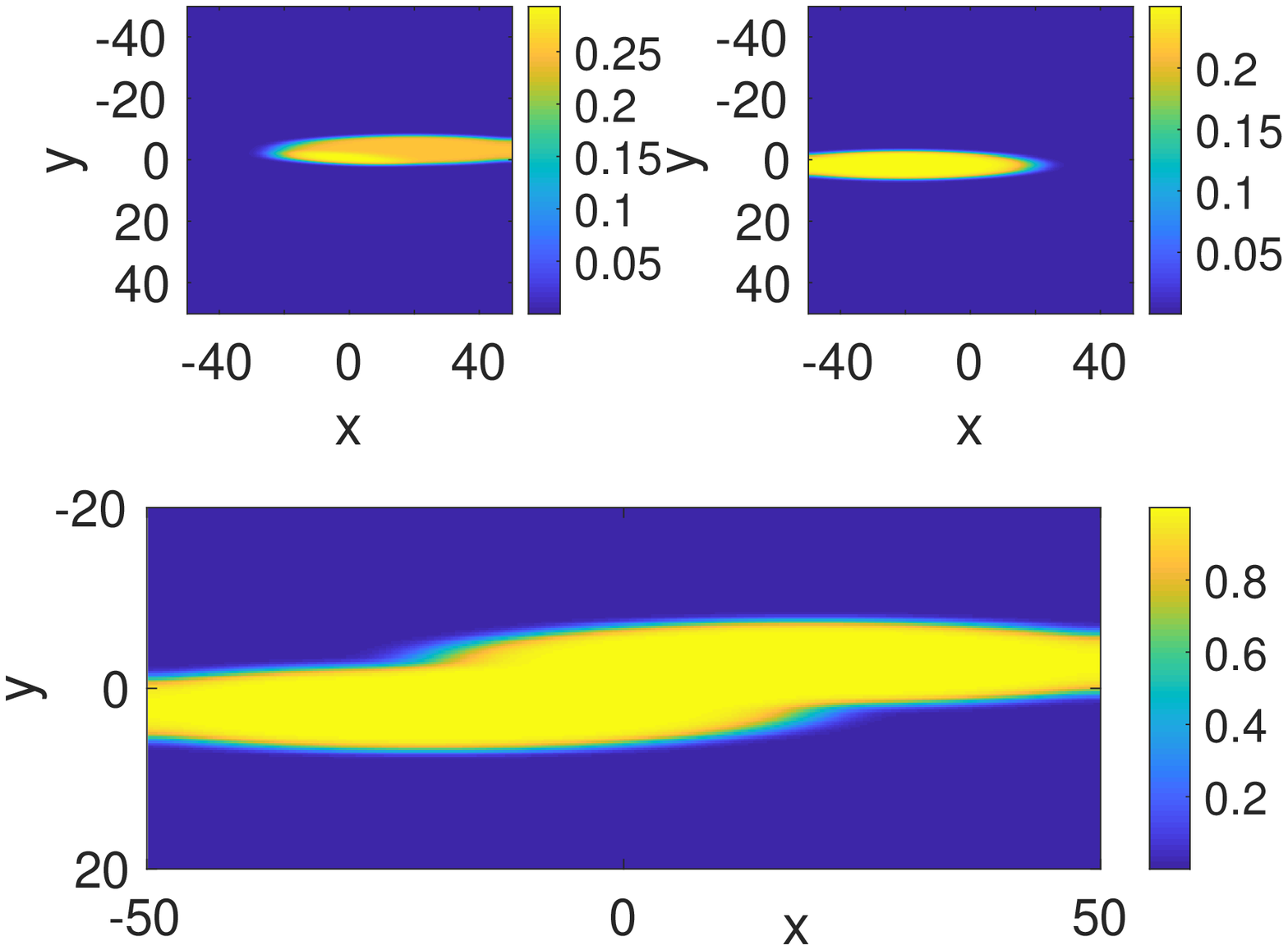}
\caption{Similar to the results of the primary tauopathy case in the
  non-collinear case of Fig.~\ref{dark_sp_1b}, but now
in the scenario of a two-species model only for the toxic components
of A$\beta$ and $\tau$P. Again only the two toxic components and the
damage variable are shown in each triplet of panels.}
\label{dark_2c_sp_2}
\end{figure}
\subsection{Reduction to FKPP}
Finally, we  considered the examples of primary tauopathy in the
context
of the FKPP-type models of Eqs.~(\ref{eqn2b})-(\ref{eqn4b}), both in the
realm of the collinear propagation of the two invasion fronts
(the toxic A$\beta$ and the toxic $\tau$P) in Fig.~\ref{dark_2c_sp_1},
as well as in that of the oblique interaction in
Fig.~\ref{dark_2c_sp_2}.
It is important to observe that in both cases the qualitative dynamics
are in close analogy to the full evolution of both the healthy and toxic
populations in Figs.~\ref{dark_sp} and~\ref{dark_sp_1b}, respectively.
Notice that in the FKPP case, we only show the two toxic species
spatial concentration contour plots at different snapshots in time,
along with the corresponding damage contour profiles.
It can be clearly seen that the qualitative correspondence persists
over the time scales shown. Nevertheless at the quantitative level,
we see that the assumption of a much higher healthy concentration
is progressively less adequate. This eventually leads to an
underestimation
of the toxic concentration of the associated proteins. Nevertheless,
the effective simplification at the level of the FKPP equations is well
suited
towards understanding the associated phenomenology in all the cases
that we have examined.

%\begin{figure}[htbp]
%\centering
%\includegraphics[scale=0.27]{}
%\includegraphics[scale=0.27]{}
%\includegraphics[scale=0.27]{}
%\caption{}
%\label{br_sols2}
%\end{figure}

\section{Conclusions}

In the present work, we have explored the evolution of toxic fronts of
proteins such as amyloid-$\beta$ and the $\tau$-protein within a
two-dimensional
terrain, i.e., the propagation of neurodegenerative waves within a two-dimensional slice.
Our formulation was based on chemical kinetics, following earlier
works
such as~\cite{g2,al1} and considering both healthy and toxic
populations
of the relevant proteins and the spatio-temporal evolution of their
concentrations. It was assumed that the healthy proteins are produced
and degraded at a given rate, and there is a conversion of the
healthy proteins into toxic ones upon interaction with a toxic ``seed''.
In the case of $\tau$P, this is further catalyzed by the presence of
toxic A$\beta$. In this setting, four equilibrium fixed points were
identified
and the heteroclinic orbits connecting them dominated the relevant
dynamics. The conditions were identified under which (parametrically)
the different fixed points exist and when all were present, their
interaction
was considered primarily in two scenarios. The first, characterized
as a primary tauopathy involved the presence of all four fixed points
(toxic fronts of A$\beta$ and $\tau$P could exist independently,
but also interact to form a toxic co-existence front). The second one,
referred to as secondary tauopathy featured no toxic $\tau$P alone,
but only in conjunction with toxic A$\beta$. It was also observed
how the two-dimensional geometry and the anisotropic diffusion can
conspire to enable these fronts to propagate along quasi-1d corridors,
but concurrently can allow the interaction of the propagating fronts
to
produce an oblique wave of toxic co-existence between the different
proteins.
Finally, a reduced model solely featuring the toxic components was
developed and it was shown that it quite adequately represents the
examples considered qualitatively, although, naturally, some of the quantitative
aspects are suitably modified.

It is particularly relevant to consider this class of models further,
both from the perspective of biological ``adequacy'' (and the
potential inclusion of suitable further biologically relevant traits)
and faithfulness and, if relevant, from the perspective of
mathematical
control and optimization. More concretely, here these models
have been illustrated from the point of view of two-dimensional
partial differential equations. However, suitable connectivity networks
exist within the brain and have been mapped~\cite{g1,g2}. Incorporating
the
associated connectivity (i.e., the adjacency matrices thereof) allows to
track relevant dynamics on a more realistic network.
This is of particular interest presently in the context of
neurodegenerative
diseases; see for a recent example of experimental observations and associated
linear modeling for Parkinson's disease the work of~\cite{henderson}.
On the other hand, it is clear that the model used here is an initial
effort
to represent the spreading of disorder when the organism is ``on the
verge''
of disease. However, it is relevant to develop a variant of this model
that may feature physiological function but may be able (upon a
suitable
``bifurcation event'') to turn to the preferentiality for disease
dynamics.
A related question is that of attempting to connect parameters
postulated herein with realistic numbers stemming from biological
experiments. Estimating production and clearance levels of these
proteins may be within reach based on recent experimental biomarker tracking
capabilities~\cite{bateman}. Other coefficients, such as those of
toxic
conversion of the proteins may be more difficult to assess but the
present model (and its distinction between different types of
tauopathies) suggests the relevance of consideration of such
experiments.

Lastly, should such a model be possible to establish on a more firm
biological basis (rather than a more phenomenological one as is done
here),
the benefits would be significant at various levels. One could
consider
how to inhibit the propagation of the fronts examined herein
and what this would require from a biological intervention
(drug administration) perspective. A controlled propagation, a slowing
down and ideally a halting of such toxic fronts would be an intriguing
target for control theory objectives applied to such infinite
dimensional
models. Enabling such a mathematical testing framework would be of
particular relevance and interest, even though recent advances (such
as
those of~\cite{linse}) suggest that this may need to be done at a more
sophisticated level, like for example that of considering
distributions
of the relevant proteins. This stems from the emerging necessity to
reduce the
flux of oligomeric (but not monomeric) forms of, e.g., A$\beta$ in order to achieve 
cognitive improvement in some of the most recent experimental
studies~\cite{linse}.
\\

\textbf{Ackowledgments.} The support for A.G. by the Engineering and
Physical Sciences Research Council of Great Britain under research
grant EP/R020205/1 is gratefully acknowledged.
This material is based upon work supported by the US National Science
Foundation under Grant  DMS-1809074
(P.G.K.). P.G.K. also acknowledges support from the Leverhulme Trust via a
Visiting Fellowship and thanks the Mathematical Institute of the University
of Oxford for its hospitality during this work.
%\section*{References}

\bibliographystyle{elsarticle-num}
%\bibliography{biblio}
%\end{document}

\end{document}